\documentclass[reprint,aps,amsmath,amssymb,nofootinbib,twoside,prd,showkeys,superscriptaddress]{revtex4-1}
\pdfoutput=1
\usepackage{verbatim}
\usepackage{comment}
\usepackage{graphicx}
\usepackage[T1]{fontenc}
\usepackage{epsfig}
\usepackage{bm}
\usepackage{tensor}
\usepackage{amssymb}
\usepackage{float}
\usepackage{amsmath}
\usepackage{subfigure}
\usepackage{dcolumn}
\usepackage[utf8]{inputenc}
\usepackage{cancel}
\usepackage[colorlinks]{hyperref}
\usepackage[usenames,dvipsnames]{color}
\hypersetup{
     breaklinks=true,
    pdfstartview={FitH},    
    colorlinks=true,       
    linkcolor=blue,          
    citecolor=red,        
    filecolor=magenta,      
    urlcolor=blue,           
    anchorcolor=green,      
    linktocpage=true
}

\def\doi{http://doi.org}

\newcommand{\onehalf}{{\textstyle\frac{1}{2}}}

\newcommand{\eref}[1]{Eq.~(\ref{#1})}

\newcommand{\HCd}{\mathcal{H}}

\def\HCdt0{\tilde{\HCd}_{0}}

\newcommand{\LCd}{\mathcal{L}}

\newcommand{\afffias}{Frankfurt Institute for Advanced Studies (FIAS), Ruth-Moufang-Strasse~1, 60438 Frankfurt am Main, Germany}
\newcommand{\affbgu}{Physics Department, Ben-Gurion University of the Negev, Beer-Sheva 84105, Israel}
\newcommand{\affcam}{DAMTP, Centre for Mathematical Sciences, University of Cambridge, Wilberforce Road, Cambridge CB3 0WA, United Kingdom}
\newcommand{\affcamast}{Kavli Institute of Cosmology (KICC), University of Cambridge, Madingley Road, Cambridge, CB3 0HA, UK}
\newcommand{\affjwg}{Fachbereich Physik, Goethe-Universit\"at, Max-von-Laue-Strasse~1, 60438~Frankfurt am Main, Germany}
\newcommand{\affgsi}{GSI Helmholtzzentrum f\"ur Schwerionenforschung GmbH, Planckstrasse~1, 64291 Darmstadt, Germany}
\newcommand{\affbahamas}{Bahamas Advanced Study Institute and Conferences, 4A Ocean Heights, Hill View Circle, Stella Maris, Long Island, The Bahamas}

\begin{document}

\title{The dark side of the torsion: Dark Energy from propagating torsion}
\author{D. Benisty}
\email{benidav@post.bgu.ac.il}
\affiliation{\afffias}\affiliation{\affcam}\affiliation{\affcamast}
\author{E. I. Guendelman}
\email{guendel@bgu.ac.il}
\affiliation{\affbgu}\affiliation{\afffias}\affiliation{\affbahamas}
\author{A. van de Venn}
\email{venn@fias.uni-frankfurt.de}
\affiliation{\afffias}
\author{D. Vasak}
\email{vasak@fias.uni-frankfurt.de}
\affiliation{\afffias}
\author{J. Struckmeier}
\email{struckmeier@fias.uni-frankfurt.de}
\affiliation{\afffias}\affiliation{\affjwg}\affiliation{\affgsi}
\author{H. Stoecker}
\email{stoecker@fias.uni-frankfurt.de}
\affiliation{\afffias}\affiliation{\affjwg}\affiliation{\affgsi}
\begin{abstract}
An extension to the Einstein-Cartan (EC) action is discussed in terms of cosmological solutions. The torsion incorporated in the EC Lagrangian is assumed to be totally anti-symmetric,  represented by a time-like axial vector $S^\mu$. The dynamics of torsion is invoked by a novel kinetic term. Here we show that this kinetic term gives rise to dark energy, while the quadratic torsion term, emanating  from the EC part, represents a stiff fluid that leads to a bouncing cosmology solution. A constraint on the bouncing solution is calculated using cosmological data from different epochs.
\end{abstract}
\maketitle
\section{Introduction} 
The nature of dark energy is a long-standing unresolved problem in current cosmology. Einstein's cosmological constant, $\Lambda$, in General Relativity (GR) has been invoked to account for the observed accelerating expansion of the universe, but failed to be understood yet in terms of field theoretical considerations. An alternative direction to account for these observations has been to re-formulate GR in terms of torsion (``Teleparallel Gravity'')  \cite{Geng:2011aj,Kofinas:2014owa,Cai:2015emx,Casalino:2020kdr,Nicosia:2020egv,ElHanafy:2020pek,Bahamonde:2020bbc,Bahamonde:2021gfp}, to try various extensions of GR, see e.g. \cite{Capozziello:2011et,Jimenez:2019woj,Capozziello:2021pcg}, or to formulate gravity as a gauge theory in analogy to Yang-Mills \cite{Hehl:1994ue,Struckmeier:2017vkf,Vasak:2019nmy,Hehl:2012pi}.

The simplest theory that incorporates torsion is the Einstein-Cartan (EC) gravity \cite{Tsoubelis:1983lnp,1995GReGr..27.1031W,Hu:2007nk,Poplawski:2011qr,Poplawski:2011jz,Magueijo:2012ug,Poplawski:2018ypb,Medina:2018rnl,Cabral:2020mst,Shaposhnikov:2020aen,Izaurieta:2020kuy,Poplawski:2020hrp,Shaposhnikov:2020frq,Kasem:2020wsp,Unger:2018oqo,Iosifidis:2021iuw,Sharma:2021fou,BorislavovVasilev:2021srn,Cabral:2021dfe,Karananas:2021zkl}. The theory is based on the Einstein-Hilbert Lagrangian of GR but in Cartan geometry which is more general then the Riemannian one. The connection then splits up into the affine portion (the Levi-Civita symbol exclusive in GR) and a tensor involving the torsion 
\begin{equation}
T\indices{^{\mu}_{\nu\sigma}} = \frac{1}{2}\left(\Gamma\indices{^{\mu}_{\nu\sigma}} - \Gamma\indices{^{\mu}_{\sigma\nu}}\right).
\end{equation}
That affine geometry with torsion has many applications \cite{1980JMP....21.1439N,German:1985tuo,deSabbata:1993ge,Carroll:1994dq,deSabbata:1994wi,Poplawski:2010kb,Skugoreva:2014ena,Gonzalez-Espinoza:2020jss,Bahamonde:2021akc,Iosifidis:2021crj,Zhang:2021ygh,Capozziello:2001mq,Guimaraes:2020drj,Bolejko:2020nbw,Morawetz:2020lea}.  Ref.~{\cite{Carroll:1994dq} investigates possible constraints on propagating (kinetic) torsion.} In this letter we investigate EC gravity with a propagating torsion that produces a dark energy term and gives rise to a bouncing solution.

 {The plan of work is as follows: Section \ref{sec:ther} introduces the theory with the novel kinetic term. Section \ref{sec:hom} presents the homogeneous solution of the theory. Section \ref{sec:num} discusses the numerical evolution and  the theory's fit to a range of data. Section \ref{sec:dis} reviews the results.}

\section{The Theory}
\label{sec:ther}
The action integral $\mathcal{S} = \int{\sqrt{-g} \left( \LCd_G + \LCd_m \right) }\, d^4x$ is based in the following ansatz for the gravity Lagrangian
\begin{equation}
\LCd_G = \frac{1}{2}\mathcal{R}\left( \Gamma\right) - \frac{4!}{2m_{}^2} \partial_{[ \mu}K_{\nu\alpha\beta]} \partial^{[\mu}K^{\nu\alpha\beta]},
\label{def:action1}
\end{equation}
that extends the Einstein-Cartan term with a kinetic term for the contortion tensor. That tensor is the deviation of the asymmetric connection from the Christoffel symbol,
\begin{equation}
    K\indices{^\nu_{\alpha\beta}} = \Gamma\indices{^\nu_{\alpha\beta}} - \left\{ \genfrac{}{}{0pt}{}{\nu}{\alpha \beta} \right\}.
\end{equation}
where $\{ \}$ is the Levi-Civita symbol:
\begin{equation}
\left\{ \genfrac{}{}{0pt}{}{\rho}{\mu \nu} \right\} = \onehalf g^{\rho\lambda} (g_{\lambda\mu,\nu}+g_{\lambda\nu,\mu}-g_{\mu\nu,\lambda}),
\end{equation}

$\LCd_m$ is the matter fields Lagrangian. Here $c = \hbar = 8\pi G = 1$, $m_{}$ is a constant with the dimension of mass, $g$ is the determinant of the metric, and $\mathcal{R}\left( \Gamma\right)$ is the Ricci scalar. The metric signature is -2. The contortion tensor, 
\begin{equation}
K_{\mu\nu\sigma} = \frac{1}{2} \left(T_{\mu\nu\sigma} + T_{\nu\sigma\mu} - T_{\sigma\mu\nu} \right),
\end{equation}
becomes identical with the torsion tensor if the latter is totally anti-symmetric which we assume in the following. 
 {
The kinetic term in the Lagrangian \eqref{def:action1}, defined in terms of the total anti-symmetric partial derivative of an anti-symmetrized contortion, corresponds to a product of external derivatives, and is thus a covariant expression. 
This is analogous to the definition of the kinetic term in electrodynamics. 
A substitution of the external derivative by the covariant derivative based on the symmetric Levi-Civita connection does not change the electromagnetic field-strength tensor since the symmetric Levi-Civita connection drops out. 
But an asymmetric connection, i.e. a connection with torsion, generates an additional term giving a torsion potential in the final action.} 

\medskip
In that case also the Ricci scalar splits up into the Levi-Civita dependent part $\bar{R}$ and the torsional part:
\begin{equation}
\begin{split}
\mathcal{R}({\Gamma}) = \bar{R} + \frac{2}{\sqrt{-g}}\partial_\lambda\left(\sqrt{-g}K_{\alpha}^{\, \lambda\alpha}\right) + K_{\rho}^{\,\alpha\lambda} K_{\lambda\alpha}^{\,\rho} + K_{\alpha}^{\,\alpha\lambda}K_{\rho\lambda}^{\,\rho}  
\\= \bar{R} + K_{\rho}^{\,\alpha\lambda} K_{\lambda\alpha}^{\,\rho}.
\end{split}
\end{equation}
 {The term $\frac{2}{\sqrt{-g}}\partial_\lambda\left(\sqrt{-g}K_{\alpha}^{\, \lambda\alpha}\right)$ is a boundary term that does not contribute to the dynamical equations. 
Since the torsion is taken to be totally anti-symmetric, the term $K_{\alpha}^{\,\alpha\lambda}K_{\rho\lambda}^{\,\rho}$ is identically zero. 
Therefore, only the "mass" ("potential") term remains.} 
With this ansatz for the contortion tensor in the Lagrangian~\eqref{def:action1} the dynamics of torsion is thus driven by that potential term acquired from the Ricci scalar, and a kinetic term from the quadratic anti-symmetric derivative:
\begin{equation}
\LCd_G = \frac{1}{2}\left(\bar{R} 
 + K\indices{^{\sigma}_{\beta\nu}} K\indices{^{\nu\beta}_{\sigma}}\right) - \frac{4!}{2m_{}^2} \partial_{[ \mu}K_{\nu\alpha\beta]} \partial^{[\mu}K^{\nu\alpha\beta]}.
\label{def:action2}
\end{equation}
A totally anti-symmetric torsion tensor can be expressed using the Levi-Civita symbol in terms of a vector field as \cite{deAndrade:1997cj,Capozziello:2001mq}
\begin{equation}
T_{\mu\alpha\beta} = \frac{1}{\sqrt{3!}}\epsilon_{\mu\alpha\beta\sigma} \tilde{S}^{\sigma} 
\label{eq:vecDef}
\end{equation}
where $\tilde{S}^{\sigma} = \sqrt{-g} S^{\sigma}$ is the vector density of weight 1. The torsion potential term in the Lagrangian is then written as:
\begin{equation}
K_{\mu\alpha\beta}\,K^{\mu\alpha\beta} = S_{\sigma} S^{\sigma},
\end{equation}
using the identity  $\frac{1}{3!}\epsilon_{\mu\alpha\beta\sigma} \epsilon^{\mu\alpha\beta\lambda} = \delta_{\,\sigma}^{\lambda} $ and realizing that $\tilde{S}_{\sigma} = g_{\mu\nu} \,\tilde{S}^{\sigma}$ has the weight -1. We re-write the kinetic term with \cite{anderson1967principles}: 
\begin{equation}
\begin{split}
4!\,\partial_{[ \mu} K_{\nu\alpha\beta]} \, 
\partial^{[\mu} K^{\nu\alpha\beta]} =  \frac{1}{g} \left(\epsilon^{\mu\nu\alpha\beta}\partial_\mu K_{\nu\alpha\beta}  \right)^2.
\end{split}
\end{equation}
With Eq.~\eqref{eq:vecDef} the kinetic term then reduces to
$\frac{1}{g} \left(\partial_\mu \tilde{S}^{\mu}  \right)^2$. The Lagrangian~\eqref{def:action2} is then re-written as
\begin{equation}
\LCd_G = \frac{1}{2} \bar{R} -  \frac{1}{2g\, m_{}^2} \left(\partial_\mu \tilde{S}^\mu  \right)^2 +  \frac{1}{2g}\tilde{S}_{\mu}\tilde{S}^{\mu},
\end{equation}
or in a fully covariant form as
\begin{equation}
\LCd_G = \frac{1}{2} \bar{R} - \frac{1}{2 m_{}^2} \left(\nabla_{\mu} S^{\mu}\right)^2 +  \frac{1}{2}S_{\mu}\,S^{\mu}.
\label{eq:lagVec}
\end{equation}
 {The term $\nabla_{\mu} S^{\mu}$} has been considered in Ref.~\cite{Carroll:1994dq} together with the term~$\partial_{[\mu} S_{\nu]}\partial^{[\mu} S^{\nu]}$. Since we investigate cosmological solutions with the ansatz  ${S}^{\mu} = \left(A(t),0,0,0\right)$, the term $\partial_{[\mu} S_{\nu]}\partial^{[\mu} S^{\nu]}$ vanishes due to the symmetries stipulated. 

\medskip
\textbf{\it{Novel "gauge" symmetry}} - The kinetic term of the torsion tensor is invariant under the ``gauge transformation''
\begin{equation}
T_{\mu\alpha\beta} \rightarrow T_{\mu\alpha\beta} + \partial_{[\mu} \Lambda_{\alpha\beta]},
\label{eq:Symet}
\end{equation}
where $\Lambda_{\alpha\beta}$ is an arbitrary anti-symmetric tensor, such that 
\begin{equation}
    \partial_{[\nu}\partial_{[\mu} \Lambda_{\alpha\beta]]} =
    \partial_{[\mu}\partial_{\nu} \Lambda_{\alpha\beta]} = 0
\end{equation}
holds. 
Then Eqs.~\eqref{eq:Symet} and \eqref{eq:vecDef} imply that   $\left(S^{\mu}_{;\mu}\right)^2$ transforms as
\begin{equation}
S^{\mu} \rightarrow S^{\mu} + \frac{1}{3!}\epsilon^{\mu\nu\alpha\beta} \partial_{\nu} \Lambda_{\alpha\beta}.
\end{equation}
That gauge symmetry is broken, though, by the potential term in the Lagrangian. 

\medskip
We wish to stress here that the kinetic term of the vector field in this Lagrangian differs from that of the Proca field, $F_{\mu\nu}F^{\mu\nu}$ \cite{Esposito-Farese:2009wbc,Heisenberg:2020xak},  and, unlike the Proca kinetic term, does contribute a density term to a homogeneous cosmological solution.
 {Moreover, the potential term has the wrong sign and implies tachyonic behavior \cite{Heisenberg:2018mxx,Heisenberg:2019ekf}, indicating that the trivial, torsion-free vacuum is a false vacuum, and that in the true vacuum torsion must have a non-trivial expectation value and give rise to non-zero vacuum energy.}

\section{Homogeneous ansatz for torsion in FLRW cosmology}
\label{sec:hom}
In this section we will show how a homogeneous and isotropic torsion density in FLRW cosmology gives rise to dark energy via the kinetic term $\left(S^{\mu}_{;\mu}\right)^2$, and to a bouncing solution arising from the potential term in the Lagrangian. That ansatz, ${S}^{\mu} = \left(A(t),0,0,0\right)$, restricting the torsion to a time-like vector, is compliant with the Copernican principle underlying the FLRW metric:
\begin{equation}
   ds^2 = -n(t)^2 dt^2+ a(t)^2 \left( dr^2 + r^2 d\Omega^2\right),
\end{equation}
where $n$ is the Lapse function and $a(t)$ is the scale factor. Applying this metric with the two degrees of freedom, the Lagrangian reduces in this "mini-superspace" to:
\begin{equation}
\begin{split}
\mathcal{L}_{MSS} = -\frac{3 a^2 \ddot{a}}{n}+\frac{3 a^2 s_0^2 \dot{a} \dot{n}}{m_{}^2 n^4}+\frac{3 a^2 \dot{a} \dot{n}}{n^2}-\frac{3 a^2 s_0 \dot{a} \dot{s}_0}{m_{}^2 n^3}\\-\frac{9 a s_0^2 \dot{a}^2}{2 m_{}^2 n^3}-\frac{3 a \dot{a}^2}{n}+\frac{a^3 s_0 \dot{n} \dot{s}_0}{m_{}^2 n^4}-\frac{a^3 s_0^2 \dot{n}^2}{2 m_{}^2 n^5}-\frac{a^3 \dot{s}_0^2}{2 m_{}^2 n^3}-\frac{a^3 s_0^2}{2 n}.   
\end{split}
\end{equation}
Now the variations w.r.t. $n$ and $a$ give:
\begin{subequations}
\begin{equation}
\rho = \frac{-2 s_0 \left(3 s_0 \dot{H}+\ddot{s}_0\right)+9 H^2 s_0^2+\dot{s}_0^2}{2 m_{}^2} \\ - \frac{s_0^2}{2} + \rho_m,
\end{equation}
\begin{equation}
p = -\frac{2 s_0 \left(3 s_0 \dot{H}+\ddot{s}_0\right)+12 H s_0 \dot{s}_0+9 H^2 s_0^2+\dot{s}_0^2}{2 m_{}^2} \\ - \frac{s_0^2}{2} + p_m,
\end{equation}
\end{subequations}
with the gauge $n = 1$. The variation w.r.t. the vector field $s_0$ gives:
\begin{equation}
3 H \dot{s}_0+\ddot{s}_0 = s_0 \left(m\mu_\mu^2-3 \dot{H}\right).
\end{equation}
Setting $A := a^3 s_0$, the density and pressure equations, and the vector field variation, become:
\begin{subequations}
\begin{equation}
\rho = \frac{-2 A \ddot{A} +6 A H \dot{A} +\dot{A}^2}{2 m_{} ^2 a^6} - \frac{A^2}{2 a^6} + \rho_m,
\end{equation}
\begin{equation} 
p = -\frac{\dot{A}^2+2 A \left(\ddot{A} -3 H \dot{A}\right)}{2 m_{}^2 a^6} - \frac{A^2}{2 a^6} + p_m,
\end{equation}
\end{subequations}
\begin{equation}
m_{}^2 A + 3H \dot{A}  = \ddot{A},
\label{eq:KGlike}
\end{equation}
Inserting the vector field variation, Eq.~\eqref{eq:KGlike}, into the density and the pressure terms, gives:
\begin{subequations} \label{denspres}
\begin{equation}
3 H^2 = \frac{1}{2m_{}^2}\frac{\dot{A}^2}{a^6} - \frac{1}{2} \frac{A^2}{a^6} + \rho_m \label{density}
\end{equation}
\begin{equation}
-3 H^2 - 2\dot{H} = -\frac{1}{2m_{}^2}\frac{\dot{A}^2}{a^6} - \frac{1}{2} \frac{A^2}{a^6} + p_m. \label{pressure}
\end{equation}
\end{subequations}
The first, kinetic term on the l.h.s. has manifestly the equation of state~$w = -1$, whereas the second, potential term behaves like a stiff fluid with~$w = 1$. These density and pressure terms are different from the quintessence model  \cite{Ratra:1987rm,Caldwell:1997ii,Zlatev:1998tr} where the kinetic term has an equations of state of $w = 1$ and the potential term has an equations of state of $w = -1$. 


\section{Numerical Solutions}
\label{sec:num}
The complete evolution of the Universe with the dynamical torsion modification  {can be analyzed via a dynamical system. For this we define  the dimensionless densities for  the dark energy and stiff fluid,}
\begin{equation}
\Omega_\Lambda = \frac{\dot{A}^2}{2m_{}^2 a^6 H_0^2}, \quad \Omega_{S} = -\frac{A^2}{2a^6 H_0^2},
\end{equation}
and re-define the Hubble parameter accordingly:
\begin{equation}
\begin{split}
\left(\frac{H(z)}{H_0}\right)^2  = E(z)^2 \\ = \Omega_\Lambda(z) + \Omega_{S}(z) + \Omega_m (1+z)^3 + \Omega_r (1+z)^4.
\label{eq:E}
\end{split}
\end{equation}
$\Omega_\Lambda$ is the dark energy density part, $\Omega_{S} $ is the bouncing energy density part,  
$\Omega_m$ is the matter energy density and $\Omega_r $ is the radiation energy density, $z $ is the red-shift, $H_0 $ is the Hubble constant, and $H(z)$ is the Hubble parameter. The 
dynamics of the torsion related fluids can, after a lengthy calculation, be re-written as an autonomous system,
\begin{equation}
\dot{\Omega}_\Lambda = 2 m  \sqrt{-\Omega_\Lambda \Omega_{S}},\quad \dot{\Omega}_S = -6 H \Omega_{S} +2 m_{}  \sqrt{-\Omega_\Lambda \Omega_{S}},
\label{eq:evol}
\end{equation}
where dot denotes time derivative.   {When both derivatives are zero the solution shows the asymptotic limit of the evolution. That solution, $\Omega_{S} = 0$ and $\Omega_\Lambda = Const$ and with the conservation of $\Omega_\Lambda + \Omega_S = 1$ we get: $\Omega_{S} = 0$  and $\Omega_\Lambda = 1$, which is a stable attractor describing a dark-energy dominated Universe. } 
\begin{figure}[t]
 	\centering
\includegraphics[width=0.49\textwidth]{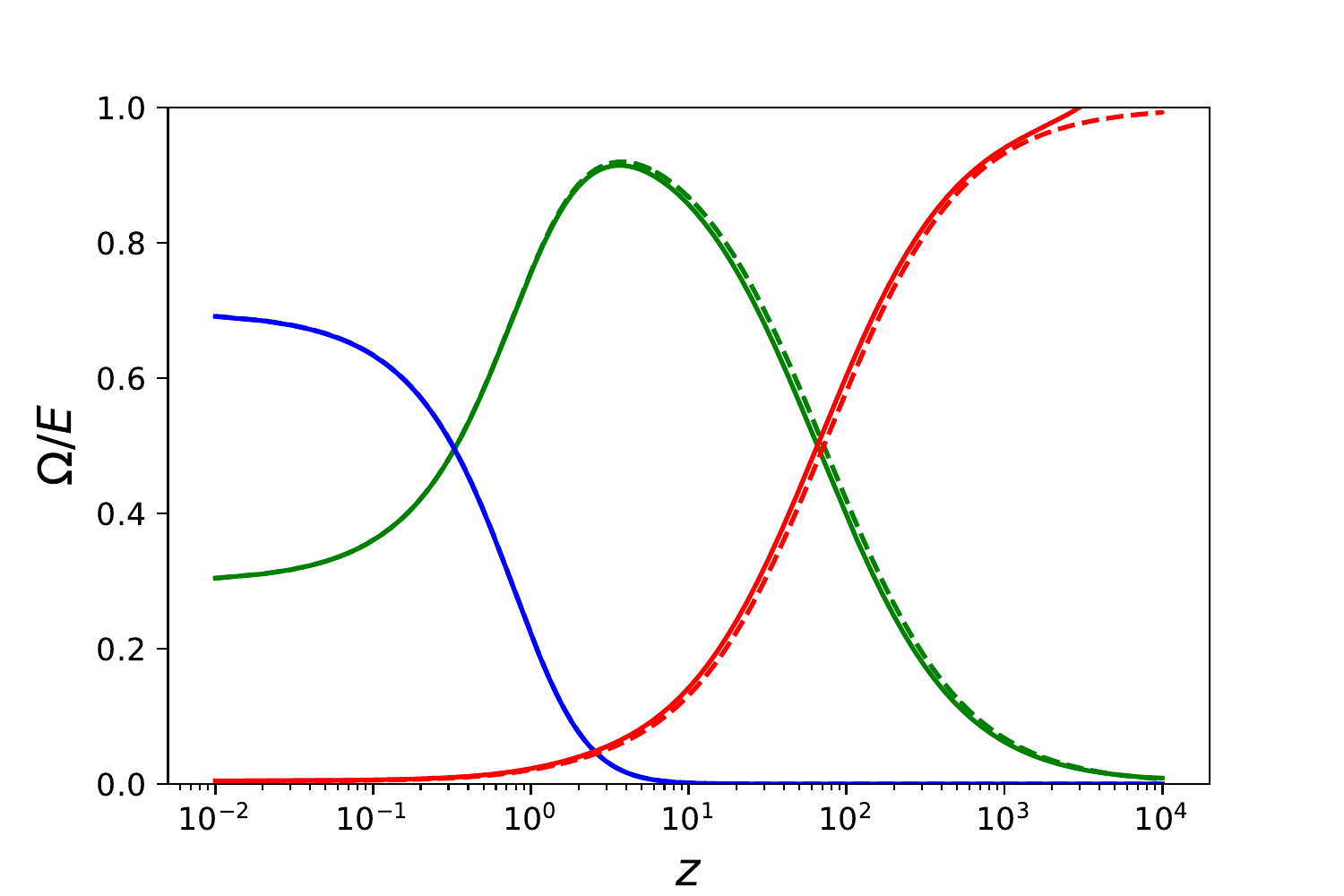}
\includegraphics[width=0.49\textwidth]{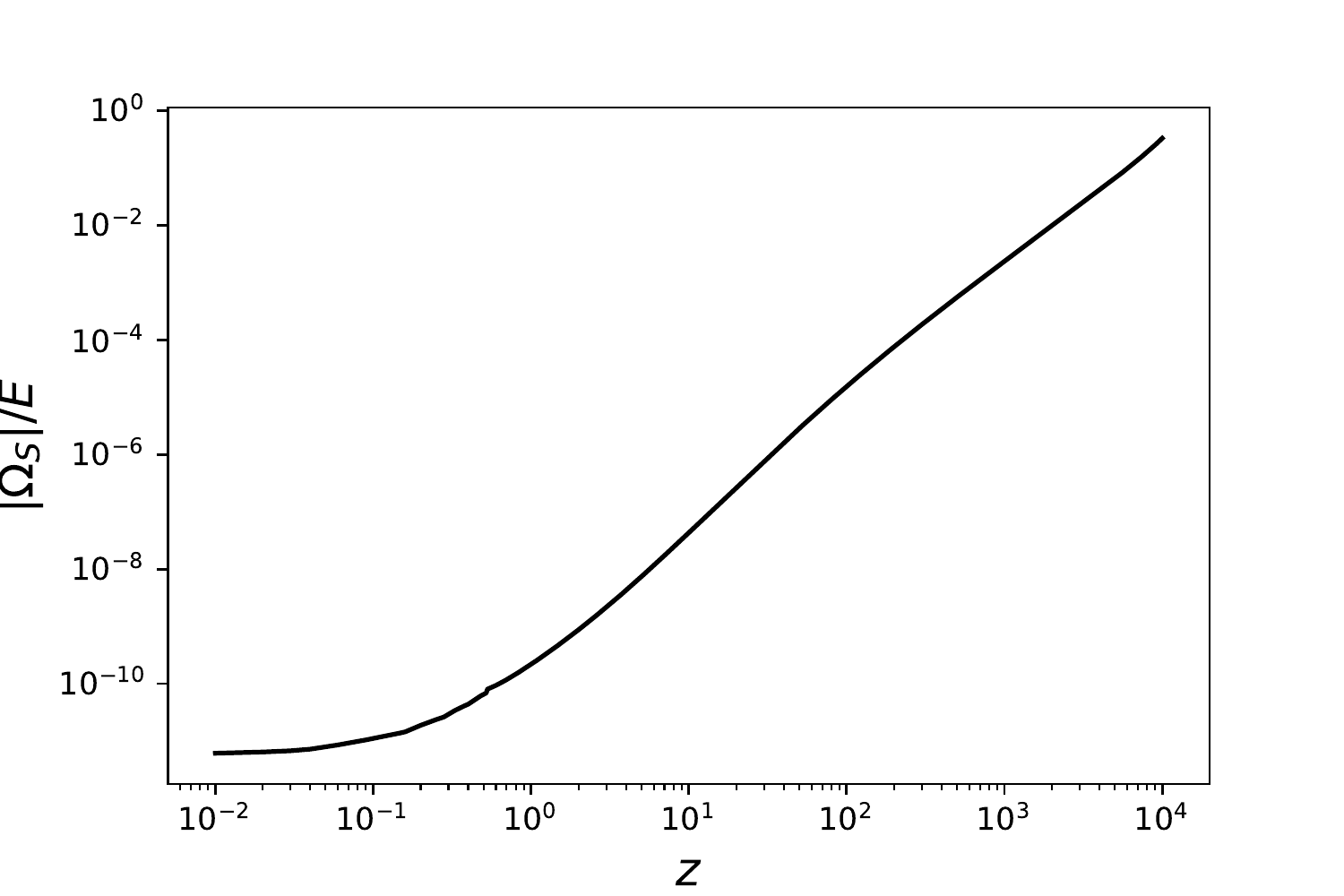}
\caption{\it{ {\textbf{Upper:} Numerical evolution of the $\Omega_m$ (green), $\Omega_\Lambda$ (blue) and $\Omega_r$ (red) for the torsional DE model. \textbf{Lower}: The $\Omega_S$ part in a logarithmic scale.}}}
 	\label{fig:Omeg} 	
\end{figure} 

\medskip
Notice furthermore that for the limit  {$m^2 \rightarrow 0$}, the kinetic dark energy part in Eqs.~\eqref{denspres} dominates. Eq.~\eqref{eq:KGlike} gives $A \sim a^3$ and the dark energy solution becomes 
\begin{equation}
\rho = \frac{\Omega_\Lambda}{2m_{}^2}H_0^2, \quad
p = -\rho,   
\end{equation}
corresponding to a dynamical Cosmological Constant. 

 {In order to track the expansion rate we rewrite the evolution equations~(\ref{eq:evol}) in terms of the red-shift:}
\begin{equation}
\begin{split}
\Omega_\Lambda'(z) &= - 2\left(\frac{m}{H_0}\right) \frac{\sqrt{-\Omega_m(z)\Omega_{S}(z)}}{(1+z)E(z)},
\\
\Omega_{S}'(z) &= 6 \frac{\Omega_{S}(z)}{z+1} + \Omega_\Lambda'(z),  
\end{split}
\label{eq:evol2}
\end{equation}
 {where $E(z)$ is defined in Eq.~(\ref{eq:E}), and comma denotes derivation w.r.t.~$z$. As an equations with dimensionless quantities, the normalized mass, $m/H_0$, is used.  }

\begin{figure*}
 	\centering
\includegraphics[width=0.9\textwidth]{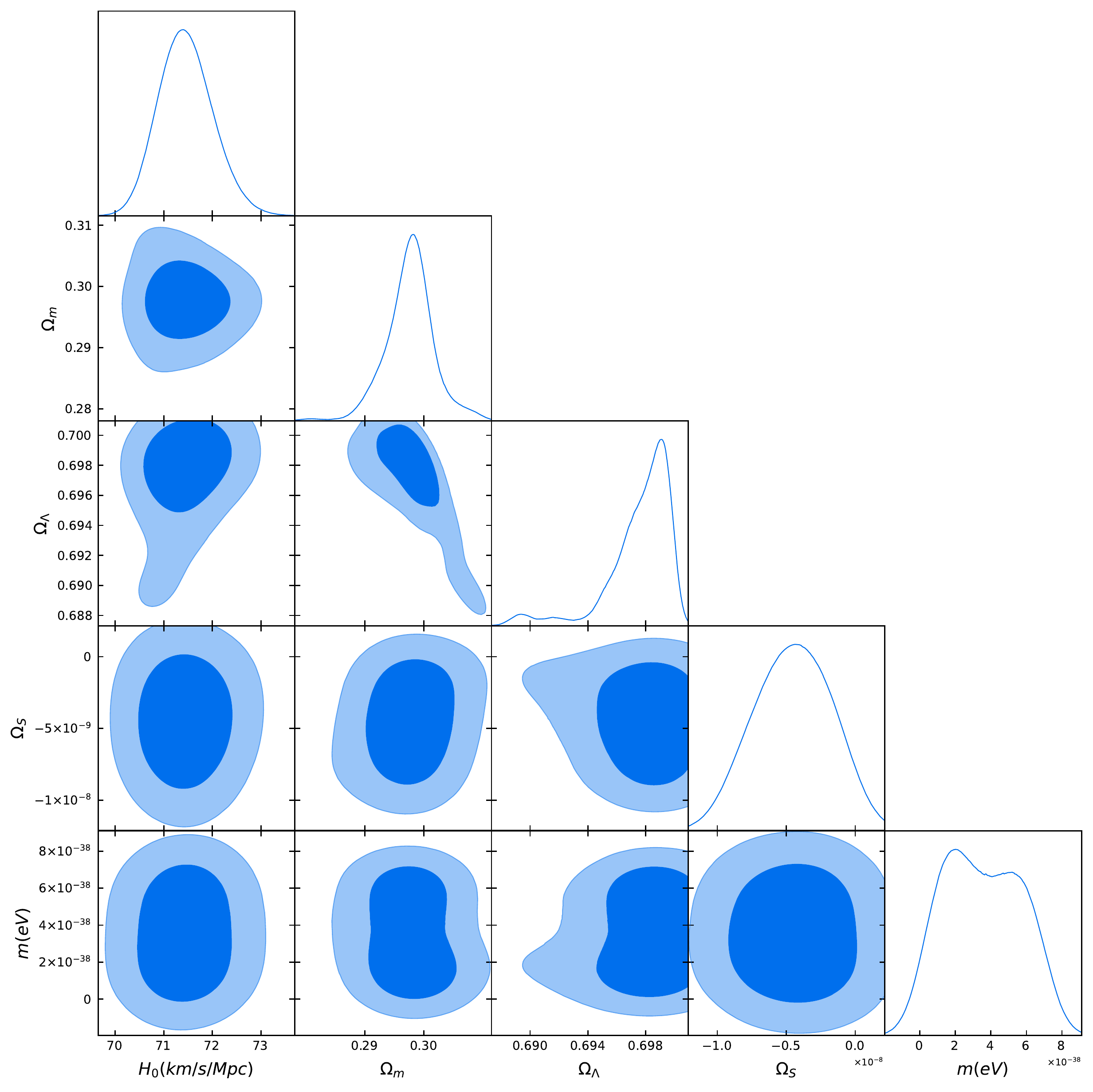}
\caption{\it{ {The posterior distributing from the BBN + CC + SC for the torsional dark energy constraint. }}}
 	\label{fig:cor}
\end{figure*}

 {For tracking the evolution of the Universe we solve Eqs.~(\ref{eq:evol2}) for the best fit values of the torsional Dark Energy model. The evolution is depicted in Fig.~\eref{fig:cor} and shows that the stiff part (with $w=1$) is dominant for the early universe (large $z$, lower panel), but 
decays in the late epoch where the kinetic term with $w=-1$ dominates (upper panel). }

\begin{table}
\centering 
\begin{tabular}{|c|c|} 
\hline \hline
parameter & Torsional DE \\
\hline \hline
 $H_0 (km/s/Mpc)$ & $71.14 \pm 0.5876$\\
$\Omega_m$ & $0.298 \pm 0.03233$ \\
$\Omega_\Lambda$ & $0.6979 \pm 0.0221$ \\
$\Omega_{S} (10^{-9})$ & $-5.811\pm 2.853$\\
$m_{} (10^{-38} eV)$ & $3.799\pm 2.063$\\
\hline\hline
\end{tabular}
	\caption{\it{ {The posterior distribution for the best fit of BBN + CC + SC.  }}}
	\label{tab:post}
\end{table}

\section{Observational Constraints} 
\label{sec:obs}
 {In order to prove the viability of the model with the torsional stiff fluid vs. observations, it is natural to first check the Big Bang Nucleosynthesis (BBN) constraint that reads:}
\begin{equation}
\left(H_{\text{TorDE}} - H_{\Lambda CDM})/H_{\Lambda CDM}\right)^2 < 10\%,
\end{equation}
 {where the Hubble parameters of this and the ${\Lambda CDM}$ models, $H_{\text{TorDE}}$ and $H_{\Lambda CDM}$, are evaluated at $z \sim 10^9$, the BBN epoch. }

 {As a second check the Cosmic Chronometers (CC) data set is used that exploits the evolution of differential ages of passive galaxies at different red-shifts to directly constrain the Hubble parameter \cite{Jimenez:2001gg}. We analyze the uncorrelated 30 CC measurements of $H(z)$, as discussed in \cite{Moresco:2012by,Moresco:2012jh,Moresco:2015cya,Moresco:2016mzx}, using Eq.~(\ref{eq:E}) for estimating the Hubble parameter for different red-shifts.}

 {As Standard Candles (SC) we use measurements of the Pantheon Type Ia supernova \cite{Scolnic:2017caz}. For each model, the pertinent model parameters are fitted to the observed $\mu _{i}^{obs}$ value by adjusting the theoretical $\mu _{i}^{th}$ value of the distance moduli,}
\begin{equation}
 \mu=m-M=5\log _{10}(D_{L})+\mu _{0}.   
\end{equation}
 {Here $m$ and $M$ are the apparent and absolute luminosity magnitudes and $\mu_{0}=5\log \left( H_{0}^{-1}/Mpc\right) +25$ is the nuisance parameter that
has been marginalized. The luminosity distance is defined as}
\begin{eqnarray}
D_L(z) &=&\frac{c}{H_{0}}(1+z)\int_{0}^{z}\frac{dz^{\ast }}{%
E(z^{\ast })}
\end{eqnarray}%
 {for $\Omega _{k}=0$ (flat space-time). 
For the SnIa data the covariance matrix is not diagonal and the distance modulus is given as the vector $\mu_{i} = \mu_{B,i}-\mathcal{M}$, where $\mu_{B,i}$ are the apparent magnitudes at maximum in the rest-frame for red-shift $z_{i}$, and $\mathcal{M}$ is treated as a universal free parameter \cite{Scolnic:2017caz} quantifying various observational uncertainties. It is apparent that the parameters $\mathcal{M}$ and $h$ are intrinsically degenerate in the context of the Pantheon data set, so we cannot extract any information regarding $H_{0}$ from SnIa data alone. }

 {Regarding the problem of likelihood maximization, we use an affine-invariant Markov Chain Monte Carlo sampler \cite{ForemanMackey:2012ig}, implemented within the open-source package $Polychord$ \cite{Handley:2015fda} with the $GetDist$ package \cite{Lewis:2019xzd}, to present the results. The prior we choose is with a uniform distribution, where $H_0 \in [50;100] km/sec/Mpc$, $\Omega_{m} \in [0;1]$, $\Omega_{\Lambda}\in[0;1 - \Omega_{m}]$, $\Omega_{S} \in [-0.1,0]$, $m_{}/H_0 \in [0,1]$. For the absolute magnitude $M$ we use a Gaussian prior $-19.28 \pm 0.0232$ taken from Ref.~\cite{Riess:2021jrx}.}

 {The posterior distribution is presented in Fig.~\ref{fig:cor} with the Table~\ref{tab:post}. The energy density of the stiff fluid is constraint to be around $10^{-4}$ in the late Universe. This value preserve also late time measurements such as CC and the SC but also for the early BBN constraint. The mass in the kinetic term is $m_{} = \left(2.924 \pm 1.933\right) \cdot 10^{-33} eV$.}

\section{Discussion} 
\label{sec:dis}
This paper extends the EC gravity by a dynamical torsion  via a kinetic term in the Lagrangian composed of totally anti-symmetric derivatives of the contortion tensor. Assuming the torsion tensor to be totally anti-symmetric, that term 
treats torsion as an additional degree of freedom equivalent to a massive vector field, $S^\mu$ . Its mass comes from the EC curvature scalar that splits up into the Ricci tensor of GR and a potential term $\sim S^\mu\,S_\mu$.

\medskip
 {A natural alternative were to replace the partial derivative in the kinetic term by a covariant derivative. In this way new torsional terms would emerge. Also the massive term could be modified, e.g. 
by adding a scalar potential $V\left(K_{\alpha\beta\gamma}K^{\alpha\beta\gamma}\right)$ to the action (\ref{eq:lagVec}). 
} generating a dynamical dark energy \cite{Copeland:2006wr,Perivolaropoulos:2021jda,Benisty:2021gde}, similarly to the three-form cosmology model \cite{Koivisto:2009ew,Koivisto:2009fb,Ngampitipan:2011se}. 
The potential 
\begin{equation}
V\left(K_{\alpha\beta\gamma}K^{\alpha\beta\gamma}\right) = - \frac{1}{2}K_{\alpha\beta\gamma}K^{\alpha\beta\gamma},    
\end{equation}
for example, cancels the bouncing term and gives rise to a Cosmological Constant that emerges from torsion. The total Lagrangian in that case reads:
\begin{equation}
\begin{split}
\LCd_G = \frac{1}{2}\mathcal{R}\left(\Gamma\right) - \frac{4!}{2m_{}^2} \partial_{[ \mu}K_{\nu\alpha\beta]} \partial^{[\mu}K^{\nu\alpha\beta]} - \frac{1}{2}K_{\alpha\beta\gamma}K^{\alpha\beta\gamma}
\\
= \frac{1}{2}\bar{R} - \frac{4!}{2m_{}^2} \partial_{[ \mu}K_{\nu\alpha\beta]} \partial^{[\mu}K^{\nu\alpha\beta]} .
\end{split}
\end{equation}
This action is equivalent to the Cosmological Constant action, but is invariant under the gauge symmetry (\ref{eq:Symet}). Possible extensions also address a scalar field $\phi$ that couples to the EC term or the torsion terms.

 {Future work might also add spin fluids in the background, or different coupling to fermionic fields. The kinetic torsion term which is the dark energy part may produce additional interactions between dark energy and matter field. Another analysis could be based on the fact that a totally anti-symmetric $(0,3)$ tensor field naturally couples to $2+1$ dimensional membranes, and this coupling produces in turn a jump in the associated $(0,4)$ field strength giving a cosmological constant. Then one can consider bubbles with different cosmological constants inside the bubble as compared to outside, and the corresponding evolution of such bubbles.}

It is important to mention that the kinetic contortion term can emerge from higher curvature theories. The Riemann-Cartan tensor can be split into the curvature part and into the torsional part via:
\begin{equation}
\begin{split}
R_{\lambda\sigma\mu\nu} \left(\Gamma\right) =
\bar{R}_{\lambda\sigma\mu\nu} &+ \bar{\nabla}_{\mu} K_{\lambda\sigma\nu} - \bar{\nabla}_{\nu} K_{\lambda\sigma\mu} \\
- K_{\lambda\beta\nu}K^{\beta}_{\, \sigma\mu}  + K_{\lambda\beta\mu}K^{\beta}_{\,\sigma\nu}.
\end{split}
\end{equation}
The terms $\partial K$ are part of the Riemann-Cartan tensor. So it is possible to obtain the term and combinations thereof (as well as additional couplings torsion-curvature) from quadratic gravity theories with $R_{\mu\nu}R^{\mu\nu}$ and/or $R_{\mu\nu\alpha\beta}R^{\mu\nu\alpha\beta}$. Gauge theories of gravity predict these terms naturally \cite{Hehl:1976kj,Ivanenko:1983fts,Ali:2009ee,Blagojevic:2012bc,Vasak:2019nmy,Benisty:2018ufz,Obukhov:2020uan,Barker:2020gcp,Benisty:2021wxi,Struckmeier:2021rst}.

\acknowledgments
D.B. acknowledges wholeheartedly the longstanding support of the Margarethe und Herbert Puschmann Stiftung. D.B gratefully acknowledge the support from the Blavatnik and the Rothschild fellowships. D.V. and A.v.d.V thank the Carl-Wilhelm Fueck Stiftung for generous support through the Walter Greiner Gesellschaft zur Foerderung der physikalischen Grundlagenforschung Frankfurt. H.St. acknowledges the Judah M. Eisenberg Professor Laureatus - Chair of the Walter Greiner Gesellschaft at Goethe Universität Frankfurt am Main. We have received partial support from European COST actions CA15117, CA16104 and CA18108 and STFC consolidated grants ST/P0006811 and ST/T0006941.

\bibliographystyle{apsrev4-1}
\bibliography{ref}

\begin{thebibliography}{84}%
\makeatletter
\providecommand \@ifxundefined [1]{%
 \@ifx{#1\undefined}
}%
\providecommand \@ifnum [1]{%
 \ifnum #1\expandafter \@firstoftwo
 \else \expandafter \@secondoftwo
 \fi
}%
\providecommand \@ifx [1]{%
 \ifx #1\expandafter \@firstoftwo
 \else \expandafter \@secondoftwo
 \fi
}%
\providecommand \natexlab [1]{#1}%
\providecommand \enquote  [1]{``#1''}%
\providecommand \bibnamefont  [1]{#1}%
\providecommand \bibfnamefont [1]{#1}%
\providecommand \citenamefont [1]{#1}%
\providecommand \href@noop [0]{\@secondoftwo}%
\providecommand \href [0]{\begingroup \@sanitize@url \@href}%
\providecommand \@href[1]{\@@startlink{#1}\@@href}%
\providecommand \@@href[1]{\endgroup#1\@@endlink}%
\providecommand \@sanitize@url [0]{\catcode `\\12\catcode `\$12\catcode
  `\&12\catcode `\#12\catcode `\^12\catcode `\_12\catcode `\%12\relax}%
\providecommand \@@startlink[1]{}%
\providecommand \@@endlink[0]{}%
\providecommand \url  [0]{\begingroup\@sanitize@url \@url }%
\providecommand \@url [1]{\endgroup\@href {#1}{\urlprefix }}%
\providecommand \urlprefix  [0]{URL }%
\providecommand \Eprint [0]{\href }%
\providecommand \doibase [0]{http://dx.doi.org/}%
\providecommand \selectlanguage [0]{\@gobble}%
\providecommand \bibinfo  [0]{\@secondoftwo}%
\providecommand \bibfield  [0]{\@secondoftwo}%
\providecommand \translation [1]{[#1]}%
\providecommand \BibitemOpen [0]{}%
\providecommand \bibitemStop [0]{}%
\providecommand \bibitemNoStop [0]{.\EOS\space}%
\providecommand \EOS [0]{\spacefactor3000\relax}%
\providecommand \BibitemShut  [1]{\csname bibitem#1\endcsname}%
\let\auto@bib@innerbib\@empty
\bibitem [{\citenamefont {Geng}\ \emph {et~al.}(2011)\citenamefont {Geng},
  \citenamefont {Lee}, \citenamefont {Saridakis},\ and\ \citenamefont
  {Wu}}]{Geng:2011aj}%
  \BibitemOpen
  \bibfield  {author} {\bibinfo {author} {\bibfnamefont {C.-Q.}\ \bibnamefont
  {Geng}}, \bibinfo {author} {\bibfnamefont {C.-C.}\ \bibnamefont {Lee}},
  \bibinfo {author} {\bibfnamefont {E.~N.}\ \bibnamefont {Saridakis}}, \ and\
  \bibinfo {author} {\bibfnamefont {Y.-P.}\ \bibnamefont {Wu}},\ }\href
  {\doibase 10.1016/j.physletb.2011.09.082} {\bibfield  {journal} {\bibinfo
  {journal} {Phys. Lett. B}\ }\textbf {\bibinfo {volume} {704}},\ \bibinfo
  {pages} {384} (\bibinfo {year} {2011})},\ \Eprint
  {http://arxiv.org/abs/1109.1092} {arXiv:1109.1092 [hep-th]} \BibitemShut
  {NoStop}%
\bibitem [{\citenamefont {Kofinas}\ and\ \citenamefont
  {Saridakis}(2014)}]{Kofinas:2014owa}%
  \BibitemOpen
  \bibfield  {author} {\bibinfo {author} {\bibfnamefont {G.}~\bibnamefont
  {Kofinas}}\ and\ \bibinfo {author} {\bibfnamefont {E.~N.}\ \bibnamefont
  {Saridakis}},\ }\href {\doibase 10.1103/PhysRevD.90.084044} {\bibfield
  {journal} {\bibinfo  {journal} {Phys. Rev. D}\ }\textbf {\bibinfo {volume}
  {90}},\ \bibinfo {pages} {084044} (\bibinfo {year} {2014})},\ \Eprint
  {http://arxiv.org/abs/1404.2249} {arXiv:1404.2249 [gr-qc]} \BibitemShut
  {NoStop}%
\bibitem [{\citenamefont {Cai}\ \emph {et~al.}(2016)\citenamefont {Cai},
  \citenamefont {Capozziello}, \citenamefont {De~Laurentis},\ and\
  \citenamefont {Saridakis}}]{Cai:2015emx}%
  \BibitemOpen
  \bibfield  {author} {\bibinfo {author} {\bibfnamefont {Y.-F.}\ \bibnamefont
  {Cai}}, \bibinfo {author} {\bibfnamefont {S.}~\bibnamefont {Capozziello}},
  \bibinfo {author} {\bibfnamefont {M.}~\bibnamefont {De~Laurentis}}, \ and\
  \bibinfo {author} {\bibfnamefont {E.~N.}\ \bibnamefont {Saridakis}},\ }\href
  {\doibase 10.1088/0034-4885/79/10/106901} {\bibfield  {journal} {\bibinfo
  {journal} {Rept. Prog. Phys.}\ }\textbf {\bibinfo {volume} {79}},\ \bibinfo
  {pages} {106901} (\bibinfo {year} {2016})},\ \Eprint
  {http://arxiv.org/abs/1511.07586} {arXiv:1511.07586 [gr-qc]} \BibitemShut
  {NoStop}%
\bibitem [{\citenamefont {Casalino}\ \emph {et~al.}(2021)\citenamefont
  {Casalino}, \citenamefont {Sanna}, \citenamefont {Sebastiani},\ and\
  \citenamefont {Zerbini}}]{Casalino:2020kdr}%
  \BibitemOpen
  \bibfield  {author} {\bibinfo {author} {\bibfnamefont {A.}~\bibnamefont
  {Casalino}}, \bibinfo {author} {\bibfnamefont {B.}~\bibnamefont {Sanna}},
  \bibinfo {author} {\bibfnamefont {L.}~\bibnamefont {Sebastiani}}, \ and\
  \bibinfo {author} {\bibfnamefont {S.}~\bibnamefont {Zerbini}},\ }\href
  {\doibase 10.1103/PhysRevD.103.023514} {\bibfield  {journal} {\bibinfo
  {journal} {Phys. Rev. D}\ }\textbf {\bibinfo {volume} {103}},\ \bibinfo
  {pages} {023514} (\bibinfo {year} {2021})},\ \Eprint
  {http://arxiv.org/abs/2010.07609} {arXiv:2010.07609 [gr-qc]} \BibitemShut
  {NoStop}%
\bibitem [{\citenamefont {Nicosia}\ \emph {et~al.}(2021)\citenamefont
  {Nicosia}, \citenamefont {Levi~Said},\ and\ \citenamefont
  {Gakis}}]{Nicosia:2020egv}%
  \BibitemOpen
  \bibfield  {author} {\bibinfo {author} {\bibfnamefont {G.-P.}\ \bibnamefont
  {Nicosia}}, \bibinfo {author} {\bibfnamefont {J.}~\bibnamefont {Levi~Said}},
  \ and\ \bibinfo {author} {\bibfnamefont {V.}~\bibnamefont {Gakis}},\ }\href
  {\doibase 10.1140/epjp/s13360-021-01133-4} {\bibfield  {journal} {\bibinfo
  {journal} {Eur. Phys. J. Plus}\ }\textbf {\bibinfo {volume} {136}},\ \bibinfo
  {pages} {191} (\bibinfo {year} {2021})},\ \Eprint
  {http://arxiv.org/abs/2012.11959} {arXiv:2012.11959 [gr-qc]} \BibitemShut
  {NoStop}%
\bibitem [{\citenamefont {El~Hanafy}\ and\ \citenamefont
  {Saridakis}(2020)}]{ElHanafy:2020pek}%
  \BibitemOpen
  \bibfield  {author} {\bibinfo {author} {\bibfnamefont {W.}~\bibnamefont
  {El~Hanafy}}\ and\ \bibinfo {author} {\bibfnamefont {E.~N.}\ \bibnamefont
  {Saridakis}},\ }\href@noop {} {\  (\bibinfo {year} {2020})},\ \Eprint
  {http://arxiv.org/abs/2011.15070} {arXiv:2011.15070 [gr-qc]} \BibitemShut
  {NoStop}%
\bibitem [{\citenamefont {Bahamonde}\ \emph {et~al.}(2020)\citenamefont
  {Bahamonde}, \citenamefont {Levi~Said},\ and\ \citenamefont
  {Zubair}}]{Bahamonde:2020bbc}%
  \BibitemOpen
  \bibfield  {author} {\bibinfo {author} {\bibfnamefont {S.}~\bibnamefont
  {Bahamonde}}, \bibinfo {author} {\bibfnamefont {J.}~\bibnamefont
  {Levi~Said}}, \ and\ \bibinfo {author} {\bibfnamefont {M.}~\bibnamefont
  {Zubair}},\ }\href {\doibase 10.1088/1475-7516/2020/10/024} {\bibfield
  {journal} {\bibinfo  {journal} {JCAP}\ }\textbf {\bibinfo {volume} {10}},\
  \bibinfo {pages} {024} (\bibinfo {year} {2020})},\ \Eprint
  {http://arxiv.org/abs/2006.06750} {arXiv:2006.06750 [gr-qc]} \BibitemShut
  {NoStop}%
\bibitem [{\citenamefont {Bahamonde}\ \emph {et~al.}(2021)\citenamefont
  {Bahamonde}, \citenamefont {Dialektopoulos}, \citenamefont
  {Escamilla-Rivera}, \citenamefont {Farrugia}, \citenamefont {Gakis},
  \citenamefont {Hendry}, \citenamefont {Hohmann}, \citenamefont {Said},
  \citenamefont {Mifsud},\ and\ \citenamefont
  {Di~Valentino}}]{Bahamonde:2021gfp}%
  \BibitemOpen
  \bibfield  {author} {\bibinfo {author} {\bibfnamefont {S.}~\bibnamefont
  {Bahamonde}}, \bibinfo {author} {\bibfnamefont {K.~F.}\ \bibnamefont
  {Dialektopoulos}}, \bibinfo {author} {\bibfnamefont {C.}~\bibnamefont
  {Escamilla-Rivera}}, \bibinfo {author} {\bibfnamefont {G.}~\bibnamefont
  {Farrugia}}, \bibinfo {author} {\bibfnamefont {V.}~\bibnamefont {Gakis}},
  \bibinfo {author} {\bibfnamefont {M.}~\bibnamefont {Hendry}}, \bibinfo
  {author} {\bibfnamefont {M.}~\bibnamefont {Hohmann}}, \bibinfo {author}
  {\bibfnamefont {J.~L.}\ \bibnamefont {Said}}, \bibinfo {author}
  {\bibfnamefont {J.}~\bibnamefont {Mifsud}}, \ and\ \bibinfo {author}
  {\bibfnamefont {E.}~\bibnamefont {Di~Valentino}},\ }\href@noop {} {\
  (\bibinfo {year} {2021})},\ \Eprint {http://arxiv.org/abs/2106.13793}
  {arXiv:2106.13793 [gr-qc]} \BibitemShut {NoStop}%
\bibitem [{\citenamefont {Capozziello}\ and\ \citenamefont
  {De~Laurentis}(2011)}]{Capozziello:2011et}%
  \BibitemOpen
  \bibfield  {author} {\bibinfo {author} {\bibfnamefont {S.}~\bibnamefont
  {Capozziello}}\ and\ \bibinfo {author} {\bibfnamefont {M.}~\bibnamefont
  {De~Laurentis}},\ }\href {\doibase 10.1016/j.physrep.2011.09.003} {\bibfield
  {journal} {\bibinfo  {journal} {Phys. Rept.}\ }\textbf {\bibinfo {volume}
  {509}},\ \bibinfo {pages} {167} (\bibinfo {year} {2011})},\ \Eprint
  {http://arxiv.org/abs/1108.6266} {arXiv:1108.6266 [gr-qc]} \BibitemShut
  {NoStop}%
\bibitem [{\citenamefont {Jim\'enez}\ \emph {et~al.}(2019)\citenamefont
  {Jim\'enez}, \citenamefont {Heisenberg},\ and\ \citenamefont
  {Koivisto}}]{Jimenez:2019woj}%
  \BibitemOpen
  \bibfield  {author} {\bibinfo {author} {\bibfnamefont {J.~B.}\ \bibnamefont
  {Jim\'enez}}, \bibinfo {author} {\bibfnamefont {L.}~\bibnamefont
  {Heisenberg}}, \ and\ \bibinfo {author} {\bibfnamefont {T.~S.}\ \bibnamefont
  {Koivisto}},\ }\href {\doibase 10.3390/universe5070173} {\bibfield  {journal}
  {\bibinfo  {journal} {Universe}\ }\textbf {\bibinfo {volume} {5}},\ \bibinfo
  {pages} {173} (\bibinfo {year} {2019})},\ \Eprint
  {http://arxiv.org/abs/1903.06830} {arXiv:1903.06830 [hep-th]} \BibitemShut
  {NoStop}%
\bibitem [{\citenamefont {Capozziello}\ \emph {et~al.}(2021)\citenamefont
  {Capozziello}, \citenamefont {Finch}, \citenamefont {Said},\ and\
  \citenamefont {Magro}}]{Capozziello:2021pcg}%
  \BibitemOpen
  \bibfield  {author} {\bibinfo {author} {\bibfnamefont {S.}~\bibnamefont
  {Capozziello}}, \bibinfo {author} {\bibfnamefont {A.}~\bibnamefont {Finch}},
  \bibinfo {author} {\bibfnamefont {J.~L.}\ \bibnamefont {Said}}, \ and\
  \bibinfo {author} {\bibfnamefont {A.}~\bibnamefont {Magro}},\ }\href@noop {}
  {\  (\bibinfo {year} {2021})},\ \Eprint {http://arxiv.org/abs/2108.03075}
  {arXiv:2108.03075 [gr-qc]} \BibitemShut {NoStop}%
\bibitem [{\citenamefont {Hehl}\ \emph {et~al.}(1995)\citenamefont {Hehl},
  \citenamefont {McCrea}, \citenamefont {Mielke},\ and\ \citenamefont
  {Ne'eman}}]{Hehl:1994ue}%
  \BibitemOpen
  \bibfield  {author} {\bibinfo {author} {\bibfnamefont {F.~W.}\ \bibnamefont
  {Hehl}}, \bibinfo {author} {\bibfnamefont {J.~D.}\ \bibnamefont {McCrea}},
  \bibinfo {author} {\bibfnamefont {E.~W.}\ \bibnamefont {Mielke}}, \ and\
  \bibinfo {author} {\bibfnamefont {Y.}~\bibnamefont {Ne'eman}},\ }\href
  {\doibase 10.1016/0370-1573(94)00111-F} {\bibfield  {journal} {\bibinfo
  {journal} {Phys. Rept.}\ }\textbf {\bibinfo {volume} {258}},\ \bibinfo
  {pages} {1} (\bibinfo {year} {1995})},\ \Eprint
  {http://arxiv.org/abs/gr-qc/9402012} {arXiv:gr-qc/9402012} \BibitemShut
  {NoStop}%
\bibitem [{\citenamefont {Struckmeier}\ \emph {et~al.}(2017)\citenamefont
  {Struckmeier}, \citenamefont {Muench}, \citenamefont {Vasak}, \citenamefont
  {Kirsch}, \citenamefont {Hanauske},\ and\ \citenamefont
  {Stoecker}}]{Struckmeier:2017vkf}%
  \BibitemOpen
  \bibfield  {author} {\bibinfo {author} {\bibfnamefont {J.}~\bibnamefont
  {Struckmeier}}, \bibinfo {author} {\bibfnamefont {J.}~\bibnamefont {Muench}},
  \bibinfo {author} {\bibfnamefont {D.}~\bibnamefont {Vasak}}, \bibinfo
  {author} {\bibfnamefont {J.}~\bibnamefont {Kirsch}}, \bibinfo {author}
  {\bibfnamefont {M.}~\bibnamefont {Hanauske}}, \ and\ \bibinfo {author}
  {\bibfnamefont {H.}~\bibnamefont {Stoecker}},\ }\href {\doibase
  10.1103/PhysRevD.95.124048} {\bibfield  {journal} {\bibinfo  {journal} {Phys.
  Rev. D}\ }\textbf {\bibinfo {volume} {95}},\ \bibinfo {pages} {124048}
  (\bibinfo {year} {2017})},\ \Eprint {http://arxiv.org/abs/1704.07246}
  {arXiv:1704.07246 [gr-qc]} \BibitemShut {NoStop}%
\bibitem [{\citenamefont {Vasak}\ \emph {et~al.}(2020)\citenamefont {Vasak},
  \citenamefont {Kirsch},\ and\ \citenamefont {Struckmeier}}]{Vasak:2019nmy}%
  \BibitemOpen
  \bibfield  {author} {\bibinfo {author} {\bibfnamefont {D.}~\bibnamefont
  {Vasak}}, \bibinfo {author} {\bibfnamefont {J.}~\bibnamefont {Kirsch}}, \
  and\ \bibinfo {author} {\bibfnamefont {J.}~\bibnamefont {Struckmeier}},\
  }\href {\doibase 10.1140/epjp/s13360-020-00415-7} {\bibfield  {journal}
  {\bibinfo  {journal} {Eur. Phys. J. Plus}\ }\textbf {\bibinfo {volume}
  {135}},\ \bibinfo {pages} {404} (\bibinfo {year} {2020})},\ \Eprint
  {http://arxiv.org/abs/1910.01088} {arXiv:1910.01088 [gr-qc]} \BibitemShut
  {NoStop}%
\bibitem [{\citenamefont {Hehl}(2017)}]{Hehl:2012pi}%
  \BibitemOpen
  \bibfield  {author} {\bibinfo {author} {\bibfnamefont {F.~W.}\ \bibnamefont
  {Hehl}},\ }\href {\doibase 10.1007/978-1-4939-3210-8_5} {\bibfield  {journal}
  {\bibinfo  {journal} {Einstein Stud.}\ }\textbf {\bibinfo {volume} {13}},\
  \bibinfo {pages} {145} (\bibinfo {year} {2017})},\ \Eprint
  {http://arxiv.org/abs/1204.3672} {arXiv:1204.3672 [gr-qc]} \BibitemShut
  {NoStop}%
\bibitem [{\citenamefont {Tsoubelis}(1983)}]{Tsoubelis:1983lnp}%
  \BibitemOpen
  \bibfield  {author} {\bibinfo {author} {\bibfnamefont {D.}~\bibnamefont
  {Tsoubelis}},\ }\href {\doibase 10.1103/PhysRevLett.51.2235} {\bibfield
  {journal} {\bibinfo  {journal} {Phys. Rev. Lett.}\ }\textbf {\bibinfo
  {volume} {51}},\ \bibinfo {pages} {2235} (\bibinfo {year}
  {1983})}\BibitemShut {NoStop}%
\bibitem [{\citenamefont {{Wolf}}(1995)}]{1995GReGr..27.1031W}%
  \BibitemOpen
  \bibfield  {author} {\bibinfo {author} {\bibfnamefont {C.}~\bibnamefont
  {{Wolf}}},\ }\href {\doibase 10.1007/BF02148646} {\bibfield  {journal}
  {\bibinfo  {journal} {General Relativity and Gravitation}\ }\textbf {\bibinfo
  {volume} {27}},\ \bibinfo {pages} {1031} (\bibinfo {year}
  {1995})}\BibitemShut {NoStop}%
\bibitem [{\citenamefont {Hu}\ and\ \citenamefont {Sawicki}(2007)}]{Hu:2007nk}%
  \BibitemOpen
  \bibfield  {author} {\bibinfo {author} {\bibfnamefont {W.}~\bibnamefont
  {Hu}}\ and\ \bibinfo {author} {\bibfnamefont {I.}~\bibnamefont {Sawicki}},\
  }\href {\doibase 10.1103/PhysRevD.76.064004} {\bibfield  {journal} {\bibinfo
  {journal} {Phys. Rev. D}\ }\textbf {\bibinfo {volume} {76}},\ \bibinfo
  {pages} {064004} (\bibinfo {year} {2007})},\ \Eprint
  {http://arxiv.org/abs/0705.1158} {arXiv:0705.1158 [astro-ph]} \BibitemShut
  {NoStop}%
\bibitem [{\citenamefont {Poplawski}(2013)}]{Poplawski:2011qr}%
  \BibitemOpen
  \bibfield  {author} {\bibinfo {author} {\bibfnamefont {N.~J.}\ \bibnamefont
  {Poplawski}},\ }\href {\doibase 10.1080/21672857.2013.11519725} {\bibfield
  {journal} {\bibinfo  {journal} {Astron. Rev.}\ }\textbf {\bibinfo {volume}
  {8}},\ \bibinfo {pages} {108} (\bibinfo {year} {2013})},\ \Eprint
  {http://arxiv.org/abs/1106.4859} {arXiv:1106.4859 [gr-qc]} \BibitemShut
  {NoStop}%
\bibitem [{\citenamefont {Poplawski}(2012)}]{Poplawski:2011jz}%
  \BibitemOpen
  \bibfield  {author} {\bibinfo {author} {\bibfnamefont {N.~J.}\ \bibnamefont
  {Poplawski}},\ }\href {\doibase 10.1103/PhysRevD.85.107502} {\bibfield
  {journal} {\bibinfo  {journal} {Phys. Rev. D}\ }\textbf {\bibinfo {volume}
  {85}},\ \bibinfo {pages} {107502} (\bibinfo {year} {2012})},\ \Eprint
  {http://arxiv.org/abs/1111.4595} {arXiv:1111.4595 [gr-qc]} \BibitemShut
  {NoStop}%
\bibitem [{\citenamefont {Magueijo}\ \emph {et~al.}(2013)\citenamefont
  {Magueijo}, \citenamefont {Zlosnik},\ and\ \citenamefont
  {Kibble}}]{Magueijo:2012ug}%
  \BibitemOpen
  \bibfield  {author} {\bibinfo {author} {\bibfnamefont {J.~a.}\ \bibnamefont
  {Magueijo}}, \bibinfo {author} {\bibfnamefont {T.~G.}\ \bibnamefont
  {Zlosnik}}, \ and\ \bibinfo {author} {\bibfnamefont {T.~W.~B.}\ \bibnamefont
  {Kibble}},\ }\href {\doibase 10.1103/PhysRevD.87.063504} {\bibfield
  {journal} {\bibinfo  {journal} {Phys. Rev. D}\ }\textbf {\bibinfo {volume}
  {87}},\ \bibinfo {pages} {063504} (\bibinfo {year} {2013})},\ \Eprint
  {http://arxiv.org/abs/1212.0585} {arXiv:1212.0585 [astro-ph.CO]} \BibitemShut
  {NoStop}%
\bibitem [{\citenamefont {Pop\l{}awski}(2018)}]{Poplawski:2018ypb}%
  \BibitemOpen
  \bibfield  {author} {\bibinfo {author} {\bibfnamefont {N.}~\bibnamefont
  {Pop\l{}awski}},\ }\href {\doibase 10.1142/S021827181847020X} {\bibfield
  {journal} {\bibinfo  {journal} {Int. J. Mod. Phys. D}\ }\textbf {\bibinfo
  {volume} {27}},\ \bibinfo {pages} {1847020} (\bibinfo {year} {2018})},\
  \Eprint {http://arxiv.org/abs/1801.08076} {arXiv:1801.08076 [physics.pop-ph]}
  \BibitemShut {NoStop}%
\bibitem [{\citenamefont {Medina}\ \emph {et~al.}(2019)\citenamefont {Medina},
  \citenamefont {Nowakowski},\ and\ \citenamefont {Batic}}]{Medina:2018rnl}%
  \BibitemOpen
  \bibfield  {author} {\bibinfo {author} {\bibfnamefont {S.~B.}\ \bibnamefont
  {Medina}}, \bibinfo {author} {\bibfnamefont {M.}~\bibnamefont {Nowakowski}},
  \ and\ \bibinfo {author} {\bibfnamefont {D.}~\bibnamefont {Batic}},\ }\href
  {\doibase 10.1016/j.aop.2018.11.002} {\bibfield  {journal} {\bibinfo
  {journal} {Annals Phys.}\ }\textbf {\bibinfo {volume} {400}},\ \bibinfo
  {pages} {64} (\bibinfo {year} {2019})},\ \Eprint
  {http://arxiv.org/abs/1812.04589} {arXiv:1812.04589 [gr-qc]} \BibitemShut
  {NoStop}%
\bibitem [{\citenamefont {Cabral}\ \emph {et~al.}(2020)\citenamefont {Cabral},
  \citenamefont {Lobo},\ and\ \citenamefont {Rubiera-Garcia}}]{Cabral:2020mst}%
  \BibitemOpen
  \bibfield  {author} {\bibinfo {author} {\bibfnamefont {F.}~\bibnamefont
  {Cabral}}, \bibinfo {author} {\bibfnamefont {F.~S.~N.}\ \bibnamefont {Lobo}},
  \ and\ \bibinfo {author} {\bibfnamefont {D.}~\bibnamefont {Rubiera-Garcia}},\
  }\href {\doibase 10.1103/PhysRevD.102.083509} {\bibfield  {journal} {\bibinfo
   {journal} {Phys. Rev. D}\ }\textbf {\bibinfo {volume} {102}},\ \bibinfo
  {pages} {083509} (\bibinfo {year} {2020})},\ \Eprint
  {http://arxiv.org/abs/2003.07463} {arXiv:2003.07463 [gr-qc]} \BibitemShut
  {NoStop}%
\bibitem [{\citenamefont {Shaposhnikov}\ \emph {et~al.}(2021)\citenamefont
  {Shaposhnikov}, \citenamefont {Shkerin}, \citenamefont {Timiryasov},\ and\
  \citenamefont {Zell}}]{Shaposhnikov:2020aen}%
  \BibitemOpen
  \bibfield  {author} {\bibinfo {author} {\bibfnamefont {M.}~\bibnamefont
  {Shaposhnikov}}, \bibinfo {author} {\bibfnamefont {A.}~\bibnamefont
  {Shkerin}}, \bibinfo {author} {\bibfnamefont {I.}~\bibnamefont {Timiryasov}},
  \ and\ \bibinfo {author} {\bibfnamefont {S.}~\bibnamefont {Zell}},\ }\href
  {\doibase 10.1103/PhysRevLett.126.161301} {\bibfield  {journal} {\bibinfo
  {journal} {Phys. Rev. Lett.}\ }\textbf {\bibinfo {volume} {126}},\ \bibinfo
  {pages} {161301} (\bibinfo {year} {2021})},\ \Eprint
  {http://arxiv.org/abs/2008.11686} {arXiv:2008.11686 [hep-ph]} \BibitemShut
  {NoStop}%
\bibitem [{\citenamefont {Izaurieta}\ \emph {et~al.}(2020)\citenamefont
  {Izaurieta}, \citenamefont {Medina}, \citenamefont {Merino}, \citenamefont
  {Salgado},\ and\ \citenamefont {Valdivia}}]{Izaurieta:2020kuy}%
  \BibitemOpen
  \bibfield  {author} {\bibinfo {author} {\bibfnamefont {F.}~\bibnamefont
  {Izaurieta}}, \bibinfo {author} {\bibfnamefont {P.}~\bibnamefont {Medina}},
  \bibinfo {author} {\bibfnamefont {N.}~\bibnamefont {Merino}}, \bibinfo
  {author} {\bibfnamefont {P.}~\bibnamefont {Salgado}}, \ and\ \bibinfo
  {author} {\bibfnamefont {O.}~\bibnamefont {Valdivia}},\ }\href {\doibase
  10.1007/JHEP10(2020)150} {\bibfield  {journal} {\bibinfo  {journal} {JHEP}\
  }\textbf {\bibinfo {volume} {10}},\ \bibinfo {pages} {150} (\bibinfo {year}
  {2020})},\ \Eprint {http://arxiv.org/abs/2007.07226} {arXiv:2007.07226
  [gr-qc]} \BibitemShut {NoStop}%
\bibitem [{\citenamefont {Pop\l{}awski}(2021)}]{Poplawski:2020hrp}%
  \BibitemOpen
  \bibfield  {author} {\bibinfo {author} {\bibfnamefont {N.}~\bibnamefont
  {Pop\l{}awski}},\ }\href {\doibase 10.1007/s10714-021-02790-7} {\bibfield
  {journal} {\bibinfo  {journal} {Gen. Rel. Grav.}\ }\textbf {\bibinfo {volume}
  {53}},\ \bibinfo {pages} {18} (\bibinfo {year} {2021})},\ \Eprint
  {http://arxiv.org/abs/2007.11556} {arXiv:2007.11556 [gr-qc]} \BibitemShut
  {NoStop}%
\bibitem [{\citenamefont {Shaposhnikov}\ \emph {et~al.}(2020)\citenamefont
  {Shaposhnikov}, \citenamefont {Shkerin}, \citenamefont {Timiryasov},\ and\
  \citenamefont {Zell}}]{Shaposhnikov:2020frq}%
  \BibitemOpen
  \bibfield  {author} {\bibinfo {author} {\bibfnamefont {M.}~\bibnamefont
  {Shaposhnikov}}, \bibinfo {author} {\bibfnamefont {A.}~\bibnamefont
  {Shkerin}}, \bibinfo {author} {\bibfnamefont {I.}~\bibnamefont {Timiryasov}},
  \ and\ \bibinfo {author} {\bibfnamefont {S.}~\bibnamefont {Zell}},\ }\href
  {\doibase 10.1007/JHEP10(2020)177} {\bibfield  {journal} {\bibinfo  {journal}
  {JHEP}\ }\textbf {\bibinfo {volume} {10}},\ \bibinfo {pages} {177} (\bibinfo
  {year} {2020})},\ \Eprint {http://arxiv.org/abs/2007.16158} {arXiv:2007.16158
  [hep-th]} \BibitemShut {NoStop}%
\bibitem [{\citenamefont {Kasem}\ and\ \citenamefont
  {Khalil}(2020)}]{Kasem:2020wsp}%
  \BibitemOpen
  \bibfield  {author} {\bibinfo {author} {\bibfnamefont {A.}~\bibnamefont
  {Kasem}}\ and\ \bibinfo {author} {\bibfnamefont {S.}~\bibnamefont {Khalil}},\
  }\href@noop {} {\  (\bibinfo {year} {2020})},\ \Eprint
  {http://arxiv.org/abs/2012.09888} {arXiv:2012.09888 [gr-qc]} \BibitemShut
  {NoStop}%
\bibitem [{\citenamefont {Unger}\ and\ \citenamefont
  {Pop\l{}awski}(2019)}]{Unger:2018oqo}%
  \BibitemOpen
  \bibfield  {author} {\bibinfo {author} {\bibfnamefont {G.}~\bibnamefont
  {Unger}}\ and\ \bibinfo {author} {\bibfnamefont {N.}~\bibnamefont
  {Pop\l{}awski}},\ }\href {\doibase 10.3847/1538-4357/aaf169} {\bibfield
  {journal} {\bibinfo  {journal} {Astrophys. J.}\ }\textbf {\bibinfo {volume}
  {870}},\ \bibinfo {pages} {78} (\bibinfo {year} {2019})},\ \Eprint
  {http://arxiv.org/abs/1808.08327} {arXiv:1808.08327 [gr-qc]} \BibitemShut
  {NoStop}%
\bibitem [{\citenamefont {Iosifidis}\ and\ \citenamefont
  {Ravera}(2021)}]{Iosifidis:2021iuw}%
  \BibitemOpen
  \bibfield  {author} {\bibinfo {author} {\bibfnamefont {D.}~\bibnamefont
  {Iosifidis}}\ and\ \bibinfo {author} {\bibfnamefont {L.}~\bibnamefont
  {Ravera}},\ }\href@noop {} {\  (\bibinfo {year} {2021})},\ \Eprint
  {http://arxiv.org/abs/2101.10339} {arXiv:2101.10339 [gr-qc]} \BibitemShut
  {NoStop}%
\bibitem [{\citenamefont {Sharma}\ and\ \citenamefont
  {Sur}(2021)}]{Sharma:2021fou}%
  \BibitemOpen
  \bibfield  {author} {\bibinfo {author} {\bibfnamefont {M.~K.}\ \bibnamefont
  {Sharma}}\ and\ \bibinfo {author} {\bibfnamefont {S.}~\bibnamefont {Sur}},\
  }\href@noop {} {\  (\bibinfo {year} {2021})},\ \Eprint
  {http://arxiv.org/abs/2102.01525} {arXiv:2102.01525 [gr-qc]} \BibitemShut
  {NoStop}%
\bibitem [{\citenamefont {Borislavov~Vasilev}\ \emph
  {et~al.}(2021)\citenamefont {Borislavov~Vasilev}, \citenamefont
  {Bouhmadi-L\'opez},\ and\ \citenamefont
  {Mart\'\i{}n-Moruno}}]{BorislavovVasilev:2021srn}%
  \BibitemOpen
  \bibfield  {author} {\bibinfo {author} {\bibfnamefont {T.}~\bibnamefont
  {Borislavov~Vasilev}}, \bibinfo {author} {\bibfnamefont {M.}~\bibnamefont
  {Bouhmadi-L\'opez}}, \ and\ \bibinfo {author} {\bibfnamefont
  {P.}~\bibnamefont {Mart\'\i{}n-Moruno}},\ }\href@noop {} {\  (\bibinfo {year}
  {2021})},\ \Eprint {http://arxiv.org/abs/2106.12050} {arXiv:2106.12050
  [gr-qc]} \BibitemShut {NoStop}%
\bibitem [{\citenamefont {Cabral}\ \emph {et~al.}(2021)\citenamefont {Cabral},
  \citenamefont {Lobo},\ and\ \citenamefont {Rubiera-Garcia}}]{Cabral:2021dfe}%
  \BibitemOpen
  \bibfield  {author} {\bibinfo {author} {\bibfnamefont {F.}~\bibnamefont
  {Cabral}}, \bibinfo {author} {\bibfnamefont {F.~S.~N.}\ \bibnamefont {Lobo}},
  \ and\ \bibinfo {author} {\bibfnamefont {D.}~\bibnamefont {Rubiera-Garcia}},\
  }\href@noop {} {\  (\bibinfo {year} {2021})},\ \Eprint
  {http://arxiv.org/abs/2102.02048} {arXiv:2102.02048 [gr-qc]} \BibitemShut
  {NoStop}%
\bibitem [{\citenamefont {Karananas}\ \emph {et~al.}(2021)\citenamefont
  {Karananas}, \citenamefont {Shaposhnikov}, \citenamefont {Shkerin},\ and\
  \citenamefont {Zell}}]{Karananas:2021zkl}%
  \BibitemOpen
  \bibfield  {author} {\bibinfo {author} {\bibfnamefont {G.~K.}\ \bibnamefont
  {Karananas}}, \bibinfo {author} {\bibfnamefont {M.}~\bibnamefont
  {Shaposhnikov}}, \bibinfo {author} {\bibfnamefont {A.}~\bibnamefont
  {Shkerin}}, \ and\ \bibinfo {author} {\bibfnamefont {S.}~\bibnamefont
  {Zell}},\ }\href@noop {} {\  (\bibinfo {year} {2021})},\ \Eprint
  {http://arxiv.org/abs/2106.13811} {arXiv:2106.13811 [hep-th]} \BibitemShut
  {NoStop}%
\bibitem [{\citenamefont {{Nieh}}(1980)}]{1980JMP....21.1439N}%
  \BibitemOpen
  \bibfield  {author} {\bibinfo {author} {\bibfnamefont {H.~T.}\ \bibnamefont
  {{Nieh}}},\ }\href {\doibase 10.1063/1.524570} {\bibfield  {journal}
  {\bibinfo  {journal} {Journal of Mathematical Physics}\ }\textbf {\bibinfo
  {volume} {21}},\ \bibinfo {pages} {1439} (\bibinfo {year}
  {1980})}\BibitemShut {NoStop}%
\bibitem [{\citenamefont {German}(1985)}]{German:1985tuo}%
  \BibitemOpen
  \bibfield  {author} {\bibinfo {author} {\bibfnamefont {G.}~\bibnamefont
  {German}},\ }\href {\doibase 10.1103/PhysRevD.32.3307} {\bibfield  {journal}
  {\bibinfo  {journal} {Phys. Rev. D}\ }\textbf {\bibinfo {volume} {32}},\
  \bibinfo {pages} {3307} (\bibinfo {year} {1985})}\BibitemShut {NoStop}%
\bibitem [{\citenamefont {de~Sabbata}(1994)}]{deSabbata:1993ge}%
  \BibitemOpen
  \bibfield  {author} {\bibinfo {author} {\bibfnamefont {V.}~\bibnamefont
  {de~Sabbata}},\ }\href@noop {} {\bibfield  {journal} {\bibinfo  {journal}
  {NATO Sci. Ser. C}\ }\textbf {\bibinfo {volume} {427}},\ \bibinfo {pages}
  {97} (\bibinfo {year} {1994})}\BibitemShut {NoStop}%
\bibitem [{\citenamefont {Carroll}\ and\ \citenamefont
  {Field}(1994)}]{Carroll:1994dq}%
  \BibitemOpen
  \bibfield  {author} {\bibinfo {author} {\bibfnamefont {S.~M.}\ \bibnamefont
  {Carroll}}\ and\ \bibinfo {author} {\bibfnamefont {G.~B.}\ \bibnamefont
  {Field}},\ }\href {\doibase 10.1103/PhysRevD.50.3867} {\bibfield  {journal}
  {\bibinfo  {journal} {Phys. Rev. D}\ }\textbf {\bibinfo {volume} {50}},\
  \bibinfo {pages} {3867} (\bibinfo {year} {1994})},\ \Eprint
  {http://arxiv.org/abs/gr-qc/9403058} {arXiv:gr-qc/9403058} \BibitemShut
  {NoStop}%
\bibitem [{\citenamefont {de~Sabbata}\ and\ \citenamefont
  {Sivaram}(1994)}]{deSabbata:1994wi}%
  \BibitemOpen
  \bibfield  {author} {\bibinfo {author} {\bibfnamefont {V.}~\bibnamefont
  {de~Sabbata}}\ and\ \bibinfo {author} {\bibfnamefont {C.}~\bibnamefont
  {Sivaram}},\ }\href@noop {} {\emph {\bibinfo {title} {{Spin and torsion in
  gravitation}}}}\ (\bibinfo {year} {1994})\BibitemShut {NoStop}%
\bibitem [{\citenamefont {Pop\l{}awski}(2010)}]{Poplawski:2010kb}%
  \BibitemOpen
  \bibfield  {author} {\bibinfo {author} {\bibfnamefont {N.~J.}\ \bibnamefont
  {Pop\l{}awski}},\ }\href {\doibase 10.1016/j.physletb.2010.09.056} {\bibfield
   {journal} {\bibinfo  {journal} {Phys. Lett. B}\ }\textbf {\bibinfo {volume}
  {694}},\ \bibinfo {pages} {181} (\bibinfo {year} {2010})},\ \bibinfo {note}
  {[Erratum: Phys.Lett.B 701, 672--672 (2011)]},\ \Eprint
  {http://arxiv.org/abs/1007.0587} {arXiv:1007.0587 [astro-ph.CO]} \BibitemShut
  {NoStop}%
\bibitem [{\citenamefont {Skugoreva}\ \emph {et~al.}(2015)\citenamefont
  {Skugoreva}, \citenamefont {Saridakis},\ and\ \citenamefont
  {Toporensky}}]{Skugoreva:2014ena}%
  \BibitemOpen
  \bibfield  {author} {\bibinfo {author} {\bibfnamefont {M.~A.}\ \bibnamefont
  {Skugoreva}}, \bibinfo {author} {\bibfnamefont {E.~N.}\ \bibnamefont
  {Saridakis}}, \ and\ \bibinfo {author} {\bibfnamefont {A.~V.}\ \bibnamefont
  {Toporensky}},\ }\href {\doibase 10.1103/PhysRevD.91.044023} {\bibfield
  {journal} {\bibinfo  {journal} {Phys. Rev. D}\ }\textbf {\bibinfo {volume}
  {91}},\ \bibinfo {pages} {044023} (\bibinfo {year} {2015})},\ \Eprint
  {http://arxiv.org/abs/1412.1502} {arXiv:1412.1502 [gr-qc]} \BibitemShut
  {NoStop}%
\bibitem [{\citenamefont {Gonzalez-Espinoza}\ and\ \citenamefont
  {Otalora}(2021)}]{Gonzalez-Espinoza:2020jss}%
  \BibitemOpen
  \bibfield  {author} {\bibinfo {author} {\bibfnamefont {M.}~\bibnamefont
  {Gonzalez-Espinoza}}\ and\ \bibinfo {author} {\bibfnamefont {G.}~\bibnamefont
  {Otalora}},\ }\href {\doibase 10.1140/epjc/s10052-021-09270-x} {\bibfield
  {journal} {\bibinfo  {journal} {Eur. Phys. J. C}\ }\textbf {\bibinfo {volume}
  {81}},\ \bibinfo {pages} {480} (\bibinfo {year} {2021})},\ \Eprint
  {http://arxiv.org/abs/2011.08377} {arXiv:2011.08377 [gr-qc]} \BibitemShut
  {NoStop}%
\bibitem [{\citenamefont {Bahamonde}\ and\ \citenamefont
  {Gigante~Valcarcel}(2021)}]{Bahamonde:2021akc}%
  \BibitemOpen
  \bibfield  {author} {\bibinfo {author} {\bibfnamefont {S.}~\bibnamefont
  {Bahamonde}}\ and\ \bibinfo {author} {\bibfnamefont {J.}~\bibnamefont
  {Gigante~Valcarcel}},\ }\href {\doibase 10.1140/epjc/s10052-021-09275-6}
  {\bibfield  {journal} {\bibinfo  {journal} {Eur. Phys. J. C}\ }\textbf
  {\bibinfo {volume} {81}},\ \bibinfo {pages} {495} (\bibinfo {year} {2021})},\
  \Eprint {http://arxiv.org/abs/2103.12036} {arXiv:2103.12036 [gr-qc]}
  \BibitemShut {NoStop}%
\bibitem [{\citenamefont {Iosifidis}(2021)}]{Iosifidis:2021crj}%
  \BibitemOpen
  \bibfield  {author} {\bibinfo {author} {\bibfnamefont {D.}~\bibnamefont
  {Iosifidis}},\ }\href@noop {} {\  (\bibinfo {year} {2021})},\ \Eprint
  {http://arxiv.org/abs/2104.10192} {arXiv:2104.10192 [gr-qc]} \BibitemShut
  {NoStop}%
\bibitem [{\citenamefont {Zhang}(2022)}]{Zhang:2021ygh}%
  \BibitemOpen
  \bibfield  {author} {\bibinfo {author} {\bibfnamefont {Z.}~\bibnamefont
  {Zhang}},\ }\href {\doibase 10.1088/1361-6382/ac38d1} {\bibfield  {journal}
  {\bibinfo  {journal} {Class. Quant. Grav.}\ }\textbf {\bibinfo {volume}
  {39}},\ \bibinfo {pages} {015003} (\bibinfo {year} {2022})},\ \Eprint
  {http://arxiv.org/abs/2112.04149} {arXiv:2112.04149 [gr-qc]} \BibitemShut
  {NoStop}%
\bibitem [{\citenamefont {Capozziello}\ \emph {et~al.}(2001)\citenamefont
  {Capozziello}, \citenamefont {Lambiase},\ and\ \citenamefont
  {Stornaiolo}}]{Capozziello:2001mq}%
  \BibitemOpen
  \bibfield  {author} {\bibinfo {author} {\bibfnamefont {S.}~\bibnamefont
  {Capozziello}}, \bibinfo {author} {\bibfnamefont {G.}~\bibnamefont
  {Lambiase}}, \ and\ \bibinfo {author} {\bibfnamefont {C.}~\bibnamefont
  {Stornaiolo}},\ }\href {\doibase
  10.1002/1521-3889(200108)10:8<713::AID-ANDP713>3.0.CO;2-2} {\bibfield
  {journal} {\bibinfo  {journal} {Annalen Phys.}\ }\textbf {\bibinfo {volume}
  {10}},\ \bibinfo {pages} {713} (\bibinfo {year} {2001})},\ \Eprint
  {http://arxiv.org/abs/gr-qc/0101038} {arXiv:gr-qc/0101038} \BibitemShut
  {NoStop}%
\bibitem [{\citenamefont {Guimar\~aes}\ \emph {et~al.}(2021)\citenamefont
  {Guimar\~aes}, \citenamefont {Lima},\ and\ \citenamefont
  {Pereira}}]{Guimaraes:2020drj}%
  \BibitemOpen
  \bibfield  {author} {\bibinfo {author} {\bibfnamefont {T.~M.}\ \bibnamefont
  {Guimar\~aes}}, \bibinfo {author} {\bibfnamefont {R.~d.~C.}\ \bibnamefont
  {Lima}}, \ and\ \bibinfo {author} {\bibfnamefont {S.~H.}\ \bibnamefont
  {Pereira}},\ }\href {\doibase 10.1140/epjc/s10052-021-09076-x} {\bibfield
  {journal} {\bibinfo  {journal} {Eur. Phys. J. C}\ }\textbf {\bibinfo {volume}
  {81}},\ \bibinfo {pages} {271} (\bibinfo {year} {2021})},\ \Eprint
  {http://arxiv.org/abs/2011.13906} {arXiv:2011.13906 [gr-qc]} \BibitemShut
  {NoStop}%
\bibitem [{\citenamefont {Bolejko}\ \emph {et~al.}(2020)\citenamefont
  {Bolejko}, \citenamefont {Cinus},\ and\ \citenamefont
  {Roukema}}]{Bolejko:2020nbw}%
  \BibitemOpen
  \bibfield  {author} {\bibinfo {author} {\bibfnamefont {K.}~\bibnamefont
  {Bolejko}}, \bibinfo {author} {\bibfnamefont {M.}~\bibnamefont {Cinus}}, \
  and\ \bibinfo {author} {\bibfnamefont {B.~F.}\ \bibnamefont {Roukema}},\
  }\href {\doibase 10.1103/PhysRevD.101.104046} {\bibfield  {journal} {\bibinfo
   {journal} {Phys. Rev. D}\ }\textbf {\bibinfo {volume} {101}},\ \bibinfo
  {pages} {104046} (\bibinfo {year} {2020})},\ \Eprint
  {http://arxiv.org/abs/2003.06528} {arXiv:2003.06528 [astro-ph.CO]}
  \BibitemShut {NoStop}%
\bibitem [{\citenamefont {Morawetz}(2021)}]{Morawetz:2020lea}%
  \BibitemOpen
  \bibfield  {author} {\bibinfo {author} {\bibfnamefont {K.}~\bibnamefont
  {Morawetz}},\ }\href {\doibase 10.1088/1361-6382/ac2417} {\bibfield
  {journal} {\bibinfo  {journal} {Class. Quant. Grav.}\ }\textbf {\bibinfo
  {volume} {38}},\ \bibinfo {pages} {205003} (\bibinfo {year} {2021})},\
  \Eprint {http://arxiv.org/abs/2010.01393} {arXiv:2010.01393 [gr-qc]}
  \BibitemShut {NoStop}%
\bibitem [{\citenamefont {de~Andrade}\ and\ \citenamefont
  {Pereira}(1999)}]{deAndrade:1997cj}%
  \BibitemOpen
  \bibfield  {author} {\bibinfo {author} {\bibfnamefont {V.~C.}\ \bibnamefont
  {de~Andrade}}\ and\ \bibinfo {author} {\bibfnamefont {J.~G.}\ \bibnamefont
  {Pereira}},\ }\href {\doibase 10.1142/S0218271899000122} {\bibfield
  {journal} {\bibinfo  {journal} {Int. J. Mod. Phys. D}\ }\textbf {\bibinfo
  {volume} {8}},\ \bibinfo {pages} {141} (\bibinfo {year} {1999})},\ \Eprint
  {http://arxiv.org/abs/gr-qc/9708051} {arXiv:gr-qc/9708051} \BibitemShut
  {NoStop}%
\bibitem [{\citenamefont {Anderson}(1967)}]{anderson1967principles}%
  \BibitemOpen
  \bibfield  {author} {\bibinfo {author} {\bibfnamefont {J.}~\bibnamefont
  {Anderson}},\ }\href {https://books.google.co.il/books?id=mAVRAAAAMAAJ}
  {\emph {\bibinfo {title} {Principles of Relativity Physics}}}\ (\bibinfo
  {publisher} {Academic Press},\ \bibinfo {year} {1967})\ pp.\ \bibinfo {pages}
  {301, 347}\BibitemShut {NoStop}%
\bibitem [{\citenamefont {Esposito-Farese}\ \emph {et~al.}(2010)\citenamefont
  {Esposito-Farese}, \citenamefont {Pitrou},\ and\ \citenamefont
  {Uzan}}]{Esposito-Farese:2009wbc}%
  \BibitemOpen
  \bibfield  {author} {\bibinfo {author} {\bibfnamefont {G.}~\bibnamefont
  {Esposito-Farese}}, \bibinfo {author} {\bibfnamefont {C.}~\bibnamefont
  {Pitrou}}, \ and\ \bibinfo {author} {\bibfnamefont {J.-P.}\ \bibnamefont
  {Uzan}},\ }\href {\doibase 10.1103/PhysRevD.81.063519} {\bibfield  {journal}
  {\bibinfo  {journal} {Phys. Rev. D}\ }\textbf {\bibinfo {volume} {81}},\
  \bibinfo {pages} {063519} (\bibinfo {year} {2010})},\ \Eprint
  {http://arxiv.org/abs/0912.0481} {arXiv:0912.0481 [gr-qc]} \BibitemShut
  {NoStop}%
\bibitem [{\citenamefont {Heisenberg}\ and\ \citenamefont
  {Villarrubia-Rojo}(2021)}]{Heisenberg:2020xak}%
  \BibitemOpen
  \bibfield  {author} {\bibinfo {author} {\bibfnamefont {L.}~\bibnamefont
  {Heisenberg}}\ and\ \bibinfo {author} {\bibfnamefont {H.}~\bibnamefont
  {Villarrubia-Rojo}},\ }\href {\doibase 10.1088/1475-7516/2021/03/032}
  {\bibfield  {journal} {\bibinfo  {journal} {JCAP}\ }\textbf {\bibinfo
  {volume} {03}},\ \bibinfo {pages} {032} (\bibinfo {year} {2021})},\ \Eprint
  {http://arxiv.org/abs/2010.00513} {arXiv:2010.00513 [astro-ph.CO]}
  \BibitemShut {NoStop}%
\bibitem [{\citenamefont {Heisenberg}\ \emph {et~al.}(2018)\citenamefont
  {Heisenberg}, \citenamefont {Kase},\ and\ \citenamefont
  {Tsujikawa}}]{Heisenberg:2018mxx}%
  \BibitemOpen
  \bibfield  {author} {\bibinfo {author} {\bibfnamefont {L.}~\bibnamefont
  {Heisenberg}}, \bibinfo {author} {\bibfnamefont {R.}~\bibnamefont {Kase}}, \
  and\ \bibinfo {author} {\bibfnamefont {S.}~\bibnamefont {Tsujikawa}},\ }\href
  {\doibase 10.1103/PhysRevD.98.024038} {\bibfield  {journal} {\bibinfo
  {journal} {Phys. Rev. D}\ }\textbf {\bibinfo {volume} {98}},\ \bibinfo
  {pages} {024038} (\bibinfo {year} {2018})},\ \Eprint
  {http://arxiv.org/abs/1805.01066} {arXiv:1805.01066 [gr-qc]} \BibitemShut
  {NoStop}%
\bibitem [{\citenamefont {Heisenberg}\ and\ \citenamefont
  {Bartelmann}(2019)}]{Heisenberg:2019ekf}%
  \BibitemOpen
  \bibfield  {author} {\bibinfo {author} {\bibfnamefont {L.}~\bibnamefont
  {Heisenberg}}\ and\ \bibinfo {author} {\bibfnamefont {M.}~\bibnamefont
  {Bartelmann}},\ }\href {\doibase 10.1016/j.physletb.2019.07.004} {\bibfield
  {journal} {\bibinfo  {journal} {Phys. Lett. B}\ }\textbf {\bibinfo {volume}
  {796}},\ \bibinfo {pages} {59} (\bibinfo {year} {2019})},\ \Eprint
  {http://arxiv.org/abs/1901.01041} {arXiv:1901.01041 [astro-ph.CO]}
  \BibitemShut {NoStop}%
\bibitem [{\citenamefont {Ratra}\ and\ \citenamefont
  {Peebles}(1988)}]{Ratra:1987rm}%
  \BibitemOpen
  \bibfield  {author} {\bibinfo {author} {\bibfnamefont {B.}~\bibnamefont
  {Ratra}}\ and\ \bibinfo {author} {\bibfnamefont {P.~J.~E.}\ \bibnamefont
  {Peebles}},\ }\href {\doibase 10.1103/PhysRevD.37.3406} {\bibfield  {journal}
  {\bibinfo  {journal} {Phys. Rev. D}\ }\textbf {\bibinfo {volume} {37}},\
  \bibinfo {pages} {3406} (\bibinfo {year} {1988})}\BibitemShut {NoStop}%
\bibitem [{\citenamefont {Caldwell}\ \emph {et~al.}(1998)\citenamefont
  {Caldwell}, \citenamefont {Dave},\ and\ \citenamefont
  {Steinhardt}}]{Caldwell:1997ii}%
  \BibitemOpen
  \bibfield  {author} {\bibinfo {author} {\bibfnamefont {R.~R.}\ \bibnamefont
  {Caldwell}}, \bibinfo {author} {\bibfnamefont {R.}~\bibnamefont {Dave}}, \
  and\ \bibinfo {author} {\bibfnamefont {P.~J.}\ \bibnamefont {Steinhardt}},\
  }\href {\doibase 10.1103/PhysRevLett.80.1582} {\bibfield  {journal} {\bibinfo
   {journal} {Phys. Rev. Lett.}\ }\textbf {\bibinfo {volume} {80}},\ \bibinfo
  {pages} {1582} (\bibinfo {year} {1998})},\ \Eprint
  {http://arxiv.org/abs/astro-ph/9708069} {arXiv:astro-ph/9708069} \BibitemShut
  {NoStop}%
\bibitem [{\citenamefont {Zlatev}\ \emph {et~al.}(1999)\citenamefont {Zlatev},
  \citenamefont {Wang},\ and\ \citenamefont {Steinhardt}}]{Zlatev:1998tr}%
  \BibitemOpen
  \bibfield  {author} {\bibinfo {author} {\bibfnamefont {I.}~\bibnamefont
  {Zlatev}}, \bibinfo {author} {\bibfnamefont {L.-M.}\ \bibnamefont {Wang}}, \
  and\ \bibinfo {author} {\bibfnamefont {P.~J.}\ \bibnamefont {Steinhardt}},\
  }\href {\doibase 10.1103/PhysRevLett.82.896} {\bibfield  {journal} {\bibinfo
  {journal} {Phys. Rev. Lett.}\ }\textbf {\bibinfo {volume} {82}},\ \bibinfo
  {pages} {896} (\bibinfo {year} {1999})},\ \Eprint
  {http://arxiv.org/abs/astro-ph/9807002} {arXiv:astro-ph/9807002} \BibitemShut
  {NoStop}%
\bibitem [{\citenamefont {Jimenez}\ and\ \citenamefont
  {Loeb}(2002)}]{Jimenez:2001gg}%
  \BibitemOpen
  \bibfield  {author} {\bibinfo {author} {\bibfnamefont {R.}~\bibnamefont
  {Jimenez}}\ and\ \bibinfo {author} {\bibfnamefont {A.}~\bibnamefont {Loeb}},\
  }\href {\doibase 10.1086/340549} {\bibfield  {journal} {\bibinfo  {journal}
  {Astrophys. J.}\ }\textbf {\bibinfo {volume} {573}},\ \bibinfo {pages} {37}
  (\bibinfo {year} {2002})},\ \Eprint {http://arxiv.org/abs/astro-ph/0106145}
  {arXiv:astro-ph/0106145} \BibitemShut {NoStop}%
\bibitem [{\citenamefont {Moresco}\ \emph
  {et~al.}(2012{\natexlab{a}})\citenamefont {Moresco}, \citenamefont {Verde},
  \citenamefont {Pozzetti}, \citenamefont {Jimenez},\ and\ \citenamefont
  {Cimatti}}]{Moresco:2012by}%
  \BibitemOpen
  \bibfield  {author} {\bibinfo {author} {\bibfnamefont {M.}~\bibnamefont
  {Moresco}}, \bibinfo {author} {\bibfnamefont {L.}~\bibnamefont {Verde}},
  \bibinfo {author} {\bibfnamefont {L.}~\bibnamefont {Pozzetti}}, \bibinfo
  {author} {\bibfnamefont {R.}~\bibnamefont {Jimenez}}, \ and\ \bibinfo
  {author} {\bibfnamefont {A.}~\bibnamefont {Cimatti}},\ }\href {\doibase
  10.1088/1475-7516/2012/07/053} {\bibfield  {journal} {\bibinfo  {journal}
  {JCAP}\ }\textbf {\bibinfo {volume} {07}},\ \bibinfo {pages} {053} (\bibinfo
  {year} {2012}{\natexlab{a}})},\ \Eprint {http://arxiv.org/abs/1201.6658}
  {arXiv:1201.6658 [astro-ph.CO]} \BibitemShut {NoStop}%
\bibitem [{\citenamefont {Moresco}\ \emph
  {et~al.}(2012{\natexlab{b}})\citenamefont {Moresco} \emph
  {et~al.}}]{Moresco:2012jh}%
  \BibitemOpen
  \bibfield  {author} {\bibinfo {author} {\bibfnamefont {M.}~\bibnamefont
  {Moresco}} \emph {et~al.},\ }\href {\doibase 10.1088/1475-7516/2012/08/006}
  {\bibfield  {journal} {\bibinfo  {journal} {JCAP}\ }\textbf {\bibinfo
  {volume} {08}},\ \bibinfo {pages} {006} (\bibinfo {year}
  {2012}{\natexlab{b}})},\ \Eprint {http://arxiv.org/abs/1201.3609}
  {arXiv:1201.3609 [astro-ph.CO]} \BibitemShut {NoStop}%
\bibitem [{\citenamefont {Moresco}(2015)}]{Moresco:2015cya}%
  \BibitemOpen
  \bibfield  {author} {\bibinfo {author} {\bibfnamefont {M.}~\bibnamefont
  {Moresco}},\ }\href {\doibase 10.1093/mnrasl/slv037} {\bibfield  {journal}
  {\bibinfo  {journal} {Mon. Not. Roy. Astron. Soc.}\ }\textbf {\bibinfo
  {volume} {450}},\ \bibinfo {pages} {L16} (\bibinfo {year} {2015})},\ \Eprint
  {http://arxiv.org/abs/1503.01116} {arXiv:1503.01116 [astro-ph.CO]}
  \BibitemShut {NoStop}%
\bibitem [{\citenamefont {Moresco}\ \emph {et~al.}(2016)\citenamefont
  {Moresco}, \citenamefont {Pozzetti}, \citenamefont {Cimatti}, \citenamefont
  {Jimenez}, \citenamefont {Maraston}, \citenamefont {Verde}, \citenamefont
  {Thomas}, \citenamefont {Citro}, \citenamefont {Tojeiro},\ and\ \citenamefont
  {Wilkinson}}]{Moresco:2016mzx}%
  \BibitemOpen
  \bibfield  {author} {\bibinfo {author} {\bibfnamefont {M.}~\bibnamefont
  {Moresco}}, \bibinfo {author} {\bibfnamefont {L.}~\bibnamefont {Pozzetti}},
  \bibinfo {author} {\bibfnamefont {A.}~\bibnamefont {Cimatti}}, \bibinfo
  {author} {\bibfnamefont {R.}~\bibnamefont {Jimenez}}, \bibinfo {author}
  {\bibfnamefont {C.}~\bibnamefont {Maraston}}, \bibinfo {author}
  {\bibfnamefont {L.}~\bibnamefont {Verde}}, \bibinfo {author} {\bibfnamefont
  {D.}~\bibnamefont {Thomas}}, \bibinfo {author} {\bibfnamefont
  {A.}~\bibnamefont {Citro}}, \bibinfo {author} {\bibfnamefont
  {R.}~\bibnamefont {Tojeiro}}, \ and\ \bibinfo {author} {\bibfnamefont
  {D.}~\bibnamefont {Wilkinson}},\ }\href {\doibase
  10.1088/1475-7516/2016/05/014} {\bibfield  {journal} {\bibinfo  {journal}
  {JCAP}\ }\textbf {\bibinfo {volume} {05}},\ \bibinfo {pages} {014} (\bibinfo
  {year} {2016})},\ \Eprint {http://arxiv.org/abs/1601.01701} {arXiv:1601.01701
  [astro-ph.CO]} \BibitemShut {NoStop}%
\bibitem [{\citenamefont {Scolnic}\ \emph {et~al.}(2018)\citenamefont {Scolnic}
  \emph {et~al.}}]{Scolnic:2017caz}%
  \BibitemOpen
  \bibfield  {author} {\bibinfo {author} {\bibfnamefont {D.~M.}\ \bibnamefont
  {Scolnic}} \emph {et~al.} (\bibinfo {collaboration} {Pan-STARRS1}),\ }\href
  {\doibase 10.3847/1538-4357/aab9bb} {\bibfield  {journal} {\bibinfo
  {journal} {Astrophys. J.}\ }\textbf {\bibinfo {volume} {859}},\ \bibinfo
  {pages} {101} (\bibinfo {year} {2018})},\ \Eprint
  {http://arxiv.org/abs/1710.00845} {arXiv:1710.00845 [astro-ph.CO]}
  \BibitemShut {NoStop}%
\bibitem [{\citenamefont {Foreman-Mackey}\ \emph {et~al.}(2013)\citenamefont
  {Foreman-Mackey}, \citenamefont {Hogg}, \citenamefont {Lang},\ and\
  \citenamefont {Goodman}}]{ForemanMackey:2012ig}%
  \BibitemOpen
  \bibfield  {author} {\bibinfo {author} {\bibfnamefont {D.}~\bibnamefont
  {Foreman-Mackey}}, \bibinfo {author} {\bibfnamefont {D.~W.}\ \bibnamefont
  {Hogg}}, \bibinfo {author} {\bibfnamefont {D.}~\bibnamefont {Lang}}, \ and\
  \bibinfo {author} {\bibfnamefont {J.}~\bibnamefont {Goodman}},\ }\href
  {\doibase 10.1086/670067} {\bibfield  {journal} {\bibinfo  {journal} {Publ.
  Astron. Soc. Pac.}\ }\textbf {\bibinfo {volume} {125}},\ \bibinfo {pages}
  {306} (\bibinfo {year} {2013})},\ \Eprint {http://arxiv.org/abs/1202.3665}
  {arXiv:1202.3665 [astro-ph.IM]} \BibitemShut {NoStop}%
\bibitem [{\citenamefont {Handley}\ \emph {et~al.}(2015)\citenamefont
  {Handley}, \citenamefont {Hobson},\ and\ \citenamefont
  {Lasenby}}]{Handley:2015fda}%
  \BibitemOpen
  \bibfield  {author} {\bibinfo {author} {\bibfnamefont {W.~J.}\ \bibnamefont
  {Handley}}, \bibinfo {author} {\bibfnamefont {M.~P.}\ \bibnamefont {Hobson}},
  \ and\ \bibinfo {author} {\bibfnamefont {A.~N.}\ \bibnamefont {Lasenby}},\
  }\href {\doibase 10.1093/mnrasl/slv047} {\bibfield  {journal} {\bibinfo
  {journal} {Mon. Not. Roy. Astron. Soc.}\ }\textbf {\bibinfo {volume} {450}},\
  \bibinfo {pages} {L61} (\bibinfo {year} {2015})},\ \Eprint
  {http://arxiv.org/abs/1502.01856} {arXiv:1502.01856 [astro-ph.CO]}
  \BibitemShut {NoStop}%
\bibitem [{\citenamefont {Lewis}(2019)}]{Lewis:2019xzd}%
  \BibitemOpen
  \bibfield  {author} {\bibinfo {author} {\bibfnamefont {A.}~\bibnamefont
  {Lewis}},\ }\href@noop {} {\  (\bibinfo {year} {2019})},\ \Eprint
  {http://arxiv.org/abs/1910.13970} {arXiv:1910.13970 [astro-ph.IM]}
  \BibitemShut {NoStop}%
\bibitem [{\citenamefont {Riess}\ \emph {et~al.}(2021)\citenamefont {Riess}
  \emph {et~al.}}]{Riess:2021jrx}%
  \BibitemOpen
  \bibfield  {author} {\bibinfo {author} {\bibfnamefont {A.~G.}\ \bibnamefont
  {Riess}} \emph {et~al.},\ }\href@noop {} {\  (\bibinfo {year} {2021})},\
  \Eprint {http://arxiv.org/abs/2112.04510} {arXiv:2112.04510 [astro-ph.CO]}
  \BibitemShut {NoStop}%
\bibitem [{\citenamefont {Copeland}\ \emph {et~al.}(2006)\citenamefont
  {Copeland}, \citenamefont {Sami},\ and\ \citenamefont
  {Tsujikawa}}]{Copeland:2006wr}%
  \BibitemOpen
  \bibfield  {author} {\bibinfo {author} {\bibfnamefont {E.~J.}\ \bibnamefont
  {Copeland}}, \bibinfo {author} {\bibfnamefont {M.}~\bibnamefont {Sami}}, \
  and\ \bibinfo {author} {\bibfnamefont {S.}~\bibnamefont {Tsujikawa}},\ }\href
  {\doibase 10.1142/S021827180600942X} {\bibfield  {journal} {\bibinfo
  {journal} {Int. J. Mod. Phys. D}\ }\textbf {\bibinfo {volume} {15}},\
  \bibinfo {pages} {1753} (\bibinfo {year} {2006})},\ \Eprint
  {http://arxiv.org/abs/hep-th/0603057} {arXiv:hep-th/0603057} \BibitemShut
  {NoStop}%
\bibitem [{\citenamefont {Perivolaropoulos}\ and\ \citenamefont
  {Skara}(2021)}]{Perivolaropoulos:2021jda}%
  \BibitemOpen
  \bibfield  {author} {\bibinfo {author} {\bibfnamefont {L.}~\bibnamefont
  {Perivolaropoulos}}\ and\ \bibinfo {author} {\bibfnamefont {F.}~\bibnamefont
  {Skara}},\ }\href@noop {} {\  (\bibinfo {year} {2021})},\ \Eprint
  {http://arxiv.org/abs/2105.05208} {arXiv:2105.05208 [astro-ph.CO]}
  \BibitemShut {NoStop}%
\bibitem [{\citenamefont {Benisty}\ and\ \citenamefont
  {Staicova}(2021)}]{Benisty:2021gde}%
  \BibitemOpen
  \bibfield  {author} {\bibinfo {author} {\bibfnamefont {D.}~\bibnamefont
  {Benisty}}\ and\ \bibinfo {author} {\bibfnamefont {D.}~\bibnamefont
  {Staicova}},\ }\href@noop {} {\  (\bibinfo {year} {2021})},\ \Eprint
  {http://arxiv.org/abs/2107.14129} {arXiv:2107.14129 [astro-ph.CO]}
  \BibitemShut {NoStop}%
\bibitem [{\citenamefont {Koivisto}\ and\ \citenamefont
  {Nunes}(2010)}]{Koivisto:2009ew}%
  \BibitemOpen
  \bibfield  {author} {\bibinfo {author} {\bibfnamefont {T.~S.}\ \bibnamefont
  {Koivisto}}\ and\ \bibinfo {author} {\bibfnamefont {N.~J.}\ \bibnamefont
  {Nunes}},\ }\href {\doibase 10.1016/j.physletb.2010.01.051} {\bibfield
  {journal} {\bibinfo  {journal} {Phys. Lett. B}\ }\textbf {\bibinfo {volume}
  {685}},\ \bibinfo {pages} {105} (\bibinfo {year} {2010})},\ \Eprint
  {http://arxiv.org/abs/0907.3883} {arXiv:0907.3883 [astro-ph.CO]} \BibitemShut
  {NoStop}%
\bibitem [{\citenamefont {Koivisto}\ and\ \citenamefont
  {Nunes}(2009)}]{Koivisto:2009fb}%
  \BibitemOpen
  \bibfield  {author} {\bibinfo {author} {\bibfnamefont {T.~S.}\ \bibnamefont
  {Koivisto}}\ and\ \bibinfo {author} {\bibfnamefont {N.~J.}\ \bibnamefont
  {Nunes}},\ }\href {\doibase 10.1103/PhysRevD.80.103509} {\bibfield  {journal}
  {\bibinfo  {journal} {Phys. Rev. D}\ }\textbf {\bibinfo {volume} {80}},\
  \bibinfo {pages} {103509} (\bibinfo {year} {2009})},\ \Eprint
  {http://arxiv.org/abs/0908.0920} {arXiv:0908.0920 [astro-ph.CO]} \BibitemShut
  {NoStop}%
\bibitem [{\citenamefont {Ngampitipan}\ and\ \citenamefont
  {Wongjun}(2011)}]{Ngampitipan:2011se}%
  \BibitemOpen
  \bibfield  {author} {\bibinfo {author} {\bibfnamefont {T.}~\bibnamefont
  {Ngampitipan}}\ and\ \bibinfo {author} {\bibfnamefont {P.}~\bibnamefont
  {Wongjun}},\ }\href {\doibase 10.1088/1475-7516/2011/11/036} {\bibfield
  {journal} {\bibinfo  {journal} {JCAP}\ }\textbf {\bibinfo {volume} {11}},\
  \bibinfo {pages} {036} (\bibinfo {year} {2011})},\ \Eprint
  {http://arxiv.org/abs/1108.0140} {arXiv:1108.0140 [hep-ph]} \BibitemShut
  {NoStop}%
\bibitem [{\citenamefont {Hehl}\ \emph {et~al.}(1976)\citenamefont {Hehl},
  \citenamefont {Von Der~Heyde}, \citenamefont {Kerlick},\ and\ \citenamefont
  {Nester}}]{Hehl:1976kj}%
  \BibitemOpen
  \bibfield  {author} {\bibinfo {author} {\bibfnamefont {F.~W.}\ \bibnamefont
  {Hehl}}, \bibinfo {author} {\bibfnamefont {P.}~\bibnamefont {Von Der~Heyde}},
  \bibinfo {author} {\bibfnamefont {G.~D.}\ \bibnamefont {Kerlick}}, \ and\
  \bibinfo {author} {\bibfnamefont {J.~M.}\ \bibnamefont {Nester}},\ }\href
  {\doibase 10.1103/RevModPhys.48.393} {\bibfield  {journal} {\bibinfo
  {journal} {Rev. Mod. Phys.}\ }\textbf {\bibinfo {volume} {48}},\ \bibinfo
  {pages} {393} (\bibinfo {year} {1976})}\BibitemShut {NoStop}%
\bibitem [{\citenamefont {Ivanenko}\ and\ \citenamefont
  {Sardanashvily}(1983)}]{Ivanenko:1983fts}%
  \BibitemOpen
  \bibfield  {author} {\bibinfo {author} {\bibfnamefont {D.}~\bibnamefont
  {Ivanenko}}\ and\ \bibinfo {author} {\bibfnamefont {G.}~\bibnamefont
  {Sardanashvily}},\ }\href {\doibase 10.1016/0370-1573(83)90046-7} {\bibfield
  {journal} {\bibinfo  {journal} {Phys. Rept.}\ }\textbf {\bibinfo {volume}
  {94}},\ \bibinfo {pages} {1} (\bibinfo {year} {1983})}\BibitemShut {NoStop}%
\bibitem [{\citenamefont {Ali}\ \emph {et~al.}(2009)\citenamefont {Ali},
  \citenamefont {Cafaro}, \citenamefont {Capozziello},\ and\ \citenamefont
  {Corda}}]{Ali:2009ee}%
  \BibitemOpen
  \bibfield  {author} {\bibinfo {author} {\bibfnamefont {S.~A.}\ \bibnamefont
  {Ali}}, \bibinfo {author} {\bibfnamefont {C.}~\bibnamefont {Cafaro}},
  \bibinfo {author} {\bibfnamefont {S.}~\bibnamefont {Capozziello}}, \ and\
  \bibinfo {author} {\bibfnamefont {C.}~\bibnamefont {Corda}},\ }\href
  {\doibase 10.1007/s10773-009-0149-0} {\bibfield  {journal} {\bibinfo
  {journal} {Int. J. Theor. Phys.}\ }\textbf {\bibinfo {volume} {48}},\
  \bibinfo {pages} {3426} (\bibinfo {year} {2009})},\ \Eprint
  {http://arxiv.org/abs/0907.0934} {arXiv:0907.0934 [gr-qc]} \BibitemShut
  {NoStop}%
\bibitem [{\citenamefont {Blagojevic}\ and\ \citenamefont
  {Hehl}(2012)}]{Blagojevic:2012bc}%
  \BibitemOpen
  \bibfield  {author} {\bibinfo {author} {\bibfnamefont {M.}~\bibnamefont
  {Blagojevic}}\ and\ \bibinfo {author} {\bibfnamefont {F.~W.}\ \bibnamefont
  {Hehl}},\ }\href@noop {} {\  (\bibinfo {year} {2012})},\ \Eprint
  {http://arxiv.org/abs/1210.3775} {arXiv:1210.3775 [gr-qc]} \BibitemShut
  {NoStop}%
\bibitem [{\citenamefont {Benisty}\ \emph {et~al.}(2018)\citenamefont
  {Benisty}, \citenamefont {Guendelman}, \citenamefont {Vasak}, \citenamefont
  {Struckmeier},\ and\ \citenamefont {Stoecker}}]{Benisty:2018ufz}%
  \BibitemOpen
  \bibfield  {author} {\bibinfo {author} {\bibfnamefont {D.}~\bibnamefont
  {Benisty}}, \bibinfo {author} {\bibfnamefont {E.~I.}\ \bibnamefont
  {Guendelman}}, \bibinfo {author} {\bibfnamefont {D.}~\bibnamefont {Vasak}},
  \bibinfo {author} {\bibfnamefont {J.}~\bibnamefont {Struckmeier}}, \ and\
  \bibinfo {author} {\bibfnamefont {H.}~\bibnamefont {Stoecker}},\ }\href
  {\doibase 10.1103/PhysRevD.98.106021} {\bibfield  {journal} {\bibinfo
  {journal} {Phys. Rev. D}\ }\textbf {\bibinfo {volume} {98}},\ \bibinfo
  {pages} {106021} (\bibinfo {year} {2018})},\ \Eprint
  {http://arxiv.org/abs/1809.10447} {arXiv:1809.10447 [gr-qc]} \BibitemShut
  {NoStop}%
\bibitem [{\citenamefont {Obukhov}\ and\ \citenamefont
  {Hehl}(2020)}]{Obukhov:2020uan}%
  \BibitemOpen
  \bibfield  {author} {\bibinfo {author} {\bibfnamefont {Y.~N.}\ \bibnamefont
  {Obukhov}}\ and\ \bibinfo {author} {\bibfnamefont {F.~W.}\ \bibnamefont
  {Hehl}},\ }\href {\doibase 10.1103/PhysRevD.102.044058} {\bibfield  {journal}
  {\bibinfo  {journal} {Phys. Rev. D}\ }\textbf {\bibinfo {volume} {102}},\
  \bibinfo {pages} {044058} (\bibinfo {year} {2020})},\ \Eprint
  {http://arxiv.org/abs/2007.00043} {arXiv:2007.00043 [gr-qc]} \BibitemShut
  {NoStop}%
\bibitem [{\citenamefont {Barker}\ \emph {et~al.}(2020)\citenamefont {Barker},
  \citenamefont {Lasenby}, \citenamefont {Hobson},\ and\ \citenamefont
  {Handley}}]{Barker:2020gcp}%
  \BibitemOpen
  \bibfield  {author} {\bibinfo {author} {\bibfnamefont {W.~E.~V.}\
  \bibnamefont {Barker}}, \bibinfo {author} {\bibfnamefont {A.~N.}\
  \bibnamefont {Lasenby}}, \bibinfo {author} {\bibfnamefont {M.~P.}\
  \bibnamefont {Hobson}}, \ and\ \bibinfo {author} {\bibfnamefont {W.~J.}\
  \bibnamefont {Handley}},\ }\href {\doibase 10.1103/PhysRevD.102.024048}
  {\bibfield  {journal} {\bibinfo  {journal} {Phys. Rev. D}\ }\textbf {\bibinfo
  {volume} {102}},\ \bibinfo {pages} {024048} (\bibinfo {year} {2020})},\
  \Eprint {http://arxiv.org/abs/2003.02690} {arXiv:2003.02690 [gr-qc]}
  \BibitemShut {NoStop}%
\bibitem [{\citenamefont {Benisty}\ \emph {et~al.}(2021)\citenamefont
  {Benisty}, \citenamefont {Vasak}, \citenamefont {Kirsch},\ and\ \citenamefont
  {Struckmeier}}]{Benisty:2021wxi}%
  \BibitemOpen
  \bibfield  {author} {\bibinfo {author} {\bibfnamefont {D.}~\bibnamefont
  {Benisty}}, \bibinfo {author} {\bibfnamefont {D.}~\bibnamefont {Vasak}},
  \bibinfo {author} {\bibfnamefont {J.}~\bibnamefont {Kirsch}}, \ and\ \bibinfo
  {author} {\bibfnamefont {J.}~\bibnamefont {Struckmeier}},\ }\href {\doibase
  10.1140/epjc/s10052-021-08924-0} {\bibfield  {journal} {\bibinfo  {journal}
  {Eur. Phys. J. C}\ }\textbf {\bibinfo {volume} {81}},\ \bibinfo {pages} {125}
  (\bibinfo {year} {2021})},\ \Eprint {http://arxiv.org/abs/2101.07566}
  {arXiv:2101.07566 [gr-qc]} \BibitemShut {NoStop}%
\bibitem [{\citenamefont {Struckmeier}\ and\ \citenamefont
  {Vasak}(2021)}]{Struckmeier:2021rst}%
  \BibitemOpen
  \bibfield  {author} {\bibinfo {author} {\bibfnamefont {J.}~\bibnamefont
  {Struckmeier}}\ and\ \bibinfo {author} {\bibfnamefont {D.}~\bibnamefont
  {Vasak}},\ }\href {\doibase 10.1002/asna.202113991} {\bibfield  {journal}
  {\bibinfo  {journal} {Astron. Nachr.}\ }\textbf {\bibinfo {volume} {342}},\
  \bibinfo {pages} {745} (\bibinfo {year} {2021})},\ \Eprint
  {http://arxiv.org/abs/2101.04467} {arXiv:2101.04467 [gr-qc]} \BibitemShut
  {NoStop}%
\end{thebibliography}%

\appendix

\end{document}